# A deep learning-accelerated data assimilation and forecasting workflow for commercial-scale geologic carbon storage

Hewei Tang[a], Pengcheng Fu[a,*], Christopher S. Sherman[a], Jize Zhang[b], Xin Ju[a], François Hamon[c], Nicholas A. Azzolina[d], Matthew Burton-Kelly[d], Joseph P. Morris[a]

[a] Atmospheric, Earth, and Energy Division, Lawrence Livermore National Laboratory, Livermore, CA 94550, United States
[b] Center for Applied Scientific Computing, Lawrence Livermore National Laboratory, Livermore, CA 94550, United States
[c] Total E&P Research and Technology, Houston, TX 77002, United States
[d] Energy & Environmental Research Center, University of North Dakota, Grand Forks, ND 58202-9018, United States

ABSTRACT

Fast assimilation of monitoring data to update forecasts of pressure buildup and carbon dioxide ($CO_2$) plume migration under geologic uncertainties is a challenging problem in geologic carbon storage. The high computational cost of data assimilation with a high-dimensional parameter space impedes fast decision-making for commercial-scale reservoir management. We propose to leverage physical understandings of porous medium flow behavior with deep learning techniques to develop a fast data assimilation-reservoir response forecasting workflow. Applying an Ensemble Smoother Multiple Data Assimilation (ES-MDA) framework, the workflow updates geologic properties and predicts reservoir performance with quantified uncertainty from pressure history and $CO_2$ plumes interpreted through seismic inversion. As the most computationally expensive component in such a workflow is reservoir simulation, we developed surrogate models to predict dynamic pressure and $CO_2$ plume extents under multi-well injection. The surrogate models employ deep convolutional neural networks, specifically, a wide residual network and a residual U-Net. The workflow is validated against a flat three-dimensional reservoir model representative of a clastic shelf depositional environment. Intelligent treatments are applied to bridge between quantities in a true-3D reservoir model and those in a single-layer reservoir model. The workflow can complete history matching and reservoir forecasting with uncertainty quantification in less than one hour on a mainstream personal workstation.

## 1. Introduction

Incorporating monitoring data to reduce the uncertainty in reservoir performance forecast is an important component for geologic carbon storage (GCS). In this paper, we consider the process of computing probability density function of a model solution conditioned on the measured observations as data assimilation or inverse problems (Evensen, 2009). The process is also known as “history matching” in petroleum reservoir engineering. In subsurface applications, data assimilation (or history matching) is characterized by large number of parameters (conceptually infinite), finite amount of data with measurement errors and nonlinear fluid flow problems (Oliver et al., 2008). A major goal of data assimilation is to integrate monitoring data into subsequent forecasts such that successive forecasts are more accurate and have quantified uncertainties. Accelerating the data assimilation process can aid rapid decision-making for reservoir management and provide critical insight to GCS project site operators. Rapid data assimilation and forecasting are some of the major goals of the U.S. Department of Energy (DOE) Science-informed Machine learning to Accelerate Real Time decision making for Carbon Storage (SMART-CS) Initiative, which aims to leverage recent advances in science-based machine learning to significantly improve the efficiency and effectiveness of commercial-scale geological carbon storage.

Data available to constrain the initial geologic description and reservoir characterization are usually limited. In GCS projects, we usually do not have the resources to acquire a large amount of data sufficient for fully constraining prior geologic descriptions of saline aquifers, although that has been occasionally done for hydrocarbon reservoirs (Chen et al., 2020; Jo et al., 2021; Ma et al., 2019; Sun et al., 2021). Moreover, our understanding or prediction of the reservoir behaviors is always uncertain. For example, uncertainty exists in assumptions used to model the process of $CO_2$ injection into deep saline aquifers, such as $CO_2$ dissolution and chemical-mechanical changes (Benson and Cole, 2008; Espinoza et al., 2018; Sun et al., 2018).

* Corresponding author.
*E-mail address:* fu4@llnl.gov (P. Fu).

The quantification and reduction of uncertainty is at the core of data assimilation. Meanwhile, decision-making requires reliable quantifications of the remaining uncertainty. Rejection sampling and Markov Chain Monte Carlo methods are rigorous Bayesian approaches for model calibration and uncertainty quantification (Chen et al., 2017; Vrugt, 2016; Wu et al., 2021). However, these methods are computationally expensive and impractical for applications in problems with a high-dimensional parameter space, which is common for subsurface reservoir models with heterogeneous reservoir porosity and permeability. Ensemble-based methods such as the ensemble Kalman filter (EnKF) and ensemble smoother (ES) are fast approximate methods for high-dimensional inversion problems (Liu and Grana, 2020a, 2020b; Ma et al., 2019). Both EnKF and ES can be viewed as gradient-based methods, where at each data assimilation step, the sensitivity matrix is estimated from an ensemble of model parameters and predictions (Emerick and Reynolds, 2013). EnKF assimilates data sequentially in time, whereas ES combines all available data, including those obtained at different times, into one assimilation step. This makes ES generally faster than EnKF but with compromised ability to match observation data (Skjervheim and Evensen, 2011). Emerick and Reynolds (2013) proposed an iterative ES approach, Ensemble Smoother with Multiple Data Assimilation (ES-MDA), which was demonstrated to have better history matching performance than EnKF with comparable computational cost. The ES-MDA algorithm has been widely applied in subsurface inverse problems such as risk management in GCS (Chen et al., 2020) and contaminant transport (Kang et al., 2021; Tso et al., 2020).

In this work, we develop a computationally efficient data assimilation and reservoir forecasting workflow for GCS in deep saline aquifers. The data assimilation workflow is developed based on the ES-MDA framework, naturally accommodating the needs for uncertainty quantification. Considering the core role of measuring and predicting $CO_2$ plume migration, we use $CO_2$ plume shapes interpreted from repeat (time-lapse) three-dimensional (3D) seismic inversion as the main monitoring data in the workflow and make predictions of future $CO_2$ plume migration patterns. The objective of the workflow is to provide rapid decision support about the areal extent and shape of the $CO_2$ plume through time, which provides important information to the GCS project site operator about the amalgamation of property rights (pore space), potential storage complex risks related to $CO_2$ migration, and the Area of Review (AOR) – the region surrounding the storage project where underground sources of drinking water may be endangered by the injection activity (US EPA Regulation 40 CFR 146.84).

Two major challenges must be overcome in the development of the workflow. The first challenge is associated with conditioning a high-dimensional parameter space with high-dimensional observation data. Reducing the dimensionalities of both parameter and data spaces through proper parameterization methods is a key aspect of the method. We refer to these two tasks as “model parameterization” and “data parameterization”, respectively. The purpose of parameterization is to represent the properties or states of interest (e.g., permeability/porosity values in all reservoir cells, and observed $CO_2$ plume shapes) with low-dimensional latent variables. New property/state distributions can be generated from randomly sampled latent variables. Principal component analysis (PCA)-based approaches are commonly applied for model parameterization, especially for Gaussian random fields (Sarma et al., 2008; Vo and Durlofsky, 2016, 2014). Recently, deep neural network-based generative models such as variational auto encoders (VAEs) and generative adversarial networks (GANs) have shown promising performance in the parameterization of non-Gaussian fields (Kang et al., 2021; Laloy et al., 2018, 2017; Mo et al., 2020). For data parameterization, Tarrahi et al. (2015) proposed to reduce the dimension of microseismic data for application in an EnKF framework based on truncated singular value decomposition (SVD) (a variant of PCA). Liu and Grana (2020a) proposed to train a deep convolutional auto encoder to reduce the dimension of seismic data. In our workflow, the observed $CO_2$ plume is a binary field (details will be introduced in Section 2), which prohibits the application of PCA approaches. We propose a polar transformation strategy for the low-dimensional representation of the $CO_2$ plumes as detailed in Section 3.3. The method proves effective for the developed workflow and is also computationally efficient.

The second challenge is the high computational cost associated with the large number of reservoir simulation runs required by the workflow. To address this challenge, we develop deep learning (DL)-based surrogate models for fast and accurate prediction of pressure and $CO_2$ plume evolutions. Deep convolutional neural networks (CNNs) have achieved significant successes in high-dimensional surrogate model development. Zhu and Zabaras (2018) proposed to treat surrogate modeling of subsurface flow problem as an image regression problem. They developed a convolutional encoder-decoder network to estimate pressure and velocity fields of single-phase steady-state flow under geologic uncertainties. In the context of multiphase dynamic flow, Zhong et al. (2019) developed a conditional convolutional generative adversarial network to predict $CO_2$ plume migration from a single injection well. Mo et al. (2019b) developed a deep convolutional encoder-decoder for $CO_2$ saturation and pressure prediction at different timesteps in an open flow domain. Wen et al. (2021) developed a residual U-Net (R-U-Net) based surrogate model for $CO_2$ plume migration in a single-well injection scenario under different permeability fields, injection durations, injection rates and injection locations. Tang et al. (2020, 2021) introduced a recurrent R-U-Net structure to predict two-dimensional (2D) and 3D dynamic pressure and saturation maps for oil production data assimilation. Their study considered multiple injection and productions wells under a constant wellbore pressure control.

Additionally, a significant gap exists in surrogate model development for multi-well $CO_2$ injection scenarios under a well-group control. Well-group control refers that multiple injection wells honor a total injection target, and therefore there is a dynamic rate allocation among different injection wells. For commercial-scale GCS with an injection rate greater than 1 million metric tons per year and a lifespan more than 10 years, multi-well injection schemes are usually necessary to achieve the necessary injectivity rate (Michael et al., 2010). DL-based surrogate models from previous works seldom focused on multi-well $CO_2$ injection schemes. We propose a two-step strategy composed of prediction of injection allocation ratios and $CO_2$ plumes, as detailed in Section 4.

We aim to develop a generic workflow that can be adapted to various GCS application scenarios. However, we specify a target application scenario as detailed in Section 2 as an illustrative example of the workflow. Section 3 presents the development of the workflow based on the ES-MDA framework. We introduce the design logic of the workflow, a geostatistical model for generating heterogeneous fields, and techniques for model and data parameterization. Section 4 presents the detailed structures of DL-based surrogate models and a comprehensive evaluation of the model performance. The results of applying the surrogate models to a data assimilation problem are presented in Section 5. In Section 6, we discuss the uncertainties being considered by the workflow and its generalizability.

## 2. Target application scenario

The workflow targets GCS in deep saline aquifers which refer to saline aquifers (generally greater than 10,000 mg/L total dissolved solids) deeper than 800 m (to ensure that $CO_2$ remains in a supercritical fluid state). Deep saline aquifers have a high global capacity for GCS (Michael et al., 2010). We chose an illustrative case from Bosshart et al. (2018), which considered the effects of depositional environments on the efficiency of GCS. The petrophysical property distributions in the models in Bosshart et al. (2018) were based on the Energy & Environmental Research Center’s (EERC) Average Global Database (AGD), which contains worldwide porosity/permeability measurements from saline aquifers.

We choose the reservoir model reflecting clastic shelf depositional environment in Bosshart et al. (2018) to be the target reference scenario for our development. The reservoir model has an aerial extent of 1034 km$^2$ (square in shape with 21-mile edge lengths) and the storage reservoir has a thickness of 91.4 m. The large size of the model makes discretization a challenging task with the need to balance the simulation run time and reasonable representation of geologic heterogeneity. Bosshart et al. (2018) determined the optimal cell size to be 152.4 × 152.4 × 3.05 m$^3$ (500 × 500 × 10 ft$^3$), yielding a discretization of 211 × 211 × 30 cells. There are 30 layers in the discretized reservoir model, and the top two layers are assumed to be cap rock (shale). The remaining layers represent the reservoir and of the models and contain two facies (i.e., high permeable and low permeable facies). A stochastic realization of the flat structural reservoir model for the clastic shelf depositional environment is illustrated in Fig. 1(a). The porosity and permeability distributions are based on the median quantities for this depositional environment. Fig. 1(b) shows the porosity-permeability correlation for the clastic shelf depositional environment from AGD. We assume the injection employs four injection wells spaced at the corners of a 10,668 m x 10,668 m square and a total injection rate of 2 million metric tons of $CO_2$ per year.

Reservoir responses of pressure and $CO_2$ saturation were simulated using CMG-GEM (CMGuser's Guide, 2019) with a Peng-Robinson equation-of-state (EOS) compositional representation and Carter-Tracy aquifer model (Carter and Tracy, 1960). Key reservoir properties are summarized in Table 1. The reservoir simulation results, including pressure distribution and $CO_2$ saturation of the reference model after five years of injection are shown in Fig. 1(c).

One of the most important metrics considered in GCS reservoir management is the lateral migration of $CO_2$ plumes. The extents of $CO_2$ plumes can be assessed by repeat (time-lapse) 3D seismic surveys (sometimes referred to as "4D seismic"). Under realistic conditions, current technologies cannot resolve (1) the vertical distribution of the $CO_2$ plume within the reservoir because the thickness of reservoir is a very small fraction of the depth of the reservoir, or (2) the spatially variable gas (mostly $CO_2$) saturation within the plumes. Therefore, to generate "synthetic observation data" of $CO_2$ migration, we first use 10% gas saturation of the CMG-GEM simulated $CO_2$ plume as the threshold value to create 3D binary $CO_2$ plume maps. This threshold is chosen based on the finding that seismic response to $CO_2$ saturation is only sensitive to changes greater than 5–10% from baseline (Roach et al., 2017). The 3D 10%-saturation plumes are illustrated as magenta surfaces in Fig. 1(c) and (d). Then we project the 3D plume onto a 2D plan-view plane as shown in Fig. 1(d). The 2D projected plumes are shown in Fig. 1(e). Though a significant amount information from the reservoir simulation is lost in this process of thresholding and dimension-collapsing, we believe the 2D $CO_2$ plume map is a realistic representation of practically measurable $CO_2$ migration patterns. When testing and demonstrating the workflow, we assume the injection project has been in operation for five years. We use the reservoir pressure history in these five years measured in one monitoring well and the "current" $CO_2$ plume map (after five years of injection) to update the prior models. The monitoring well is located at the center of the reservoir and perforated at the middle reservoir layer. We then use the calibrated model to predict the $CO_2$ plume map as well as pressure evolution in the monitoring well five years in the future.

**Table 1**
Reservoir properties used in CMG-GEM simulation.

| Parameters | Value |
|---|---|
| Reservoir dimension | 32,156.4 $m$ × 32,156.4 $m$ × 91.4m |
| Grid division | 211 × 211 × 30 |
| Field injection target | 2 million metric tons/per year |
| Compositional fluid model | Peng-Robinson equation of state (Peng and Robinson, 1976) |
| $CO_2$ solubility in brine | Harvey (1996) |
| Relative permeability | High permeable sandstone and low perm clay/shale (Bennion and Bachu, 2007) |

## 3. Data assimilation workflow

### 3.1. Design philosophy

We intend to reproduce a real-world data assimilation and forecasting problem, where there is neither a reference reservoir model nor a comprehensive physical model that can capture all the details in the subsurface. The main objective of the workflow is to predict $CO_2$ plume migration and reservoir pressure buildup in the future based on existing observations. Such predictions are not only critical metrics from a reservoir management perspective, but they are capable of being validated when additional observations are made in the future. The workflow does not intend to predict quantities (such as temperature or pressure field distribution) that are not “assimilated” to constrain the model, although it is in principle straightforward to incorporate these data if they are practically measurable.

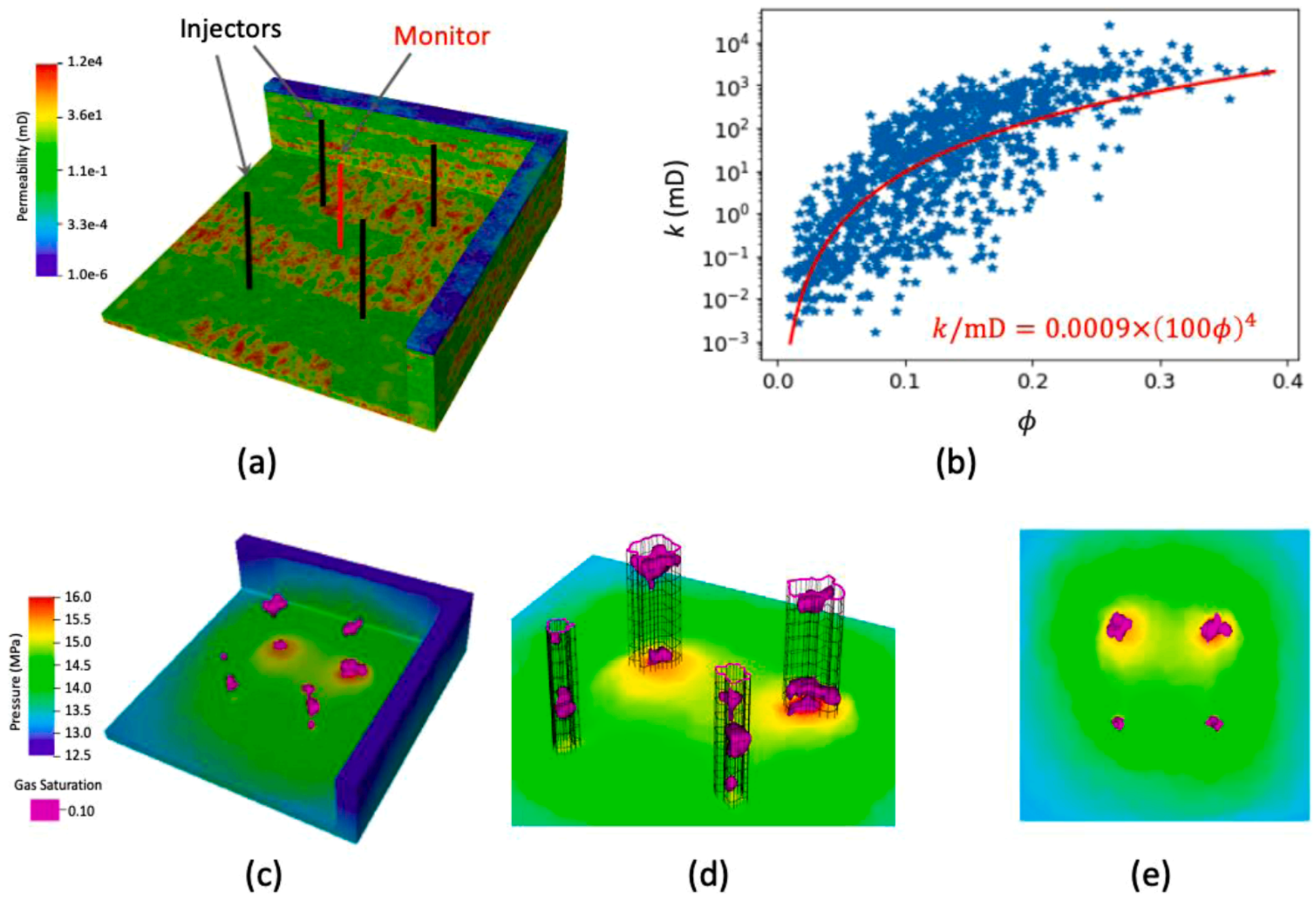

**Fig. 1.** (a) Reference reservoir model conceptualized by Bosshart et al. (2018), (b) porosity-permeability correlation of clastic shelf depositional environment given by AGD (Department of Energy National Energy Technology Laboratory, 2017), (c) reservoir simulation results showing pore pressure distribution (MPa) and $CO_2$ saturation (greater than 10%) of the reference model after five years of injection, (d) projection of the 3D plume onto a 2D plan-view plane and (e) 2D plan-view of the projected 3D plumes. Note that the vertical dimension of the reservoir model and simulation results is exaggerated by 100 times to visualize the vertical structures. Reservoir pressure is visualized in the background of (c) through (e) using the blue-to-red color ramp (For interpretation of the references to color in this figure legend, the reader is referred to the web version of this article.).

Based on the above design philosophy, we develop a data assimilation workflow as shown in Fig. 2. The workflow follows the ES-MDA inverse modeling framework proposed by Emerick and Reynolds (2013). Assuming that the workflow is blind to the reference reservoir model, the initial reservoir models are generated based on fractal distribution. PCA is applied to represent the reservoir models in a low-dimensional variable $m$. Saturation plumes and monitoring well pressure are estimated from deep learning-based surrogate models and compared with the measurements. A polar transformation technique is applied to reduce the dimensionality of saturation plumes. $m$ is then updated based on data mismatch and covariance matrixes calculated in the ES-MDA algorithm. In the following sections, we will introduce each part of the workflow in details.

### 3.2. ES-MDA framework

Given the high-dimensional parameter space, there are likely a wide range of reservoir models able to reasonably match the observation data. It is necessary to find multiple history-matched models for quantifying the uncertainties in reservoir performance predictions (Oliver and Chen, 2011). ES-MDA as an efficient ensemble-based method is well suited for this purpose. The ES-MDA algorithm updates model parameters for multiple steps with perturbed observation data. For model parameters $\mathbf{m} \in \mathbb{R}^{N_c}$ and observation data $\mathbf{d} \in \mathbb{R}^{N_d}$, the update of the model parameters of the $j^{\text{th}}$ stochastic realization in each ES-MDA iteration follows:

$$\mathbf{m}_j^l = \mathbf{m}_j^{l-1} + \mathbf{C}_{MD}^{l-1}\left(\mathbf{C}_{DD}^{l-1} + \alpha_{l-1}\mathbf{C}_D\right)^{-1}\left(\mathbf{d}_{uc,j}^{l-1} - \mathbf{d}_j^{l-1}\right) \tag{1}$$

$j = 1, 2, \ldots, N_e$, where $N_e$ is the number of stochastic realizations in the ensemble. Superscript $l$ and $l-1$ denote the current data assimilation step and the previous step, respectively. $\mathbf{d}_j^{l-1}$ is the predicted data from the underlying reservoir model with parameter $\mathbf{m}_j^{l-1}$. $\mathbf{d}_{uc,j}^{l-1}$ is perturbed observation data, namely:

$$\mathbf{d}_{uc,j}^{l-1} = \mathbf{d} + \sqrt{\alpha_{l-1}}\mathbf{C}_D^{1/2}\mathbf{z}_d^{l-1} \tag{2}$$

where $\mathbf{z}_d^{l-1} \sim N(0, \mathbf{I}_{N_d})$ and $\alpha_{l-1}$ is the inflation coefficient. $\alpha_{l-1}$ commonly takes the value of the empirically determined number of data assimilation steps ($N_a$) following the recommendation of Emerick and Reynolds (2013). $\mathbf{C}_D \in \mathbb{R}^{N_d \times N_d}$ is the covariance matrix of observed data measurement error. $\mathbf{C}_{MD}^{l-1} \in \mathbb{R}^{N_c \times N_d}$ is the cross-covariance matrix between the model parameters and the model predicted data, namely

$$\mathbf{C}_{MD}^{l-1} = \frac{1}{N_e - 1}\sum_{j=1}^{N_e}\left(\mathbf{m}_j^{l-1} - \overline{\mathbf{m}}^{l-1}\right)\left(\mathbf{d}_j^{l-1} - \overline{\mathbf{d}}^{l-1}\right)^T \tag{3}$$

$\mathbf{C}_{DD}^{l-1} \in \mathbb{R}^{N_d \times N_d}$ is the auto-covariance matrix of the predicted data,

$$\mathbf{C}_{DD}^{l-1} = \frac{1}{N_e - 1}\sum_{j=1}^{N_e}\left(\mathbf{d}_j^{l-1} - \overline{\mathbf{d}}^{l-1}\right)\left(\mathbf{d}_j^{l-1} - \overline{\mathbf{d}}^{l-1}\right)^T \tag{4}$$

Initial ensemble design is critical in ensemble-based methods (Jafarpour and McLaughlin, 2009). In the next section, we will introduce the geostatistical model for generating the initial heterogeneous permeability fields.

### 3.3. Initial reservoir models

The reservoir models underlying the workflow need to be able to generate observation data in the same space as real observation data, which are generated from simulating pressure and $CO_2$ saturation in response to $CO_2$ injection within the reference reservoir model. Considering the flat structure of the target reservoir as well as the collapsed $CO_2$ plume map, we hypothesize that a properly formulated single-layer 3D reservoir model, with some intelligent treatments, is likely to meet this need. The single-layer configuration means that the reservoir is not discretized in the vertical direction. However, it is still a 3D model and can naturally accommodate depth/thickness variations of the reservoir, although such variations are not included in the selected reference reservoir scenario.

Although the variogram models of permeability and porosity distributions for the reference reservoir model are available, we do not use such information in this work. Instead, we assume that only generic prior geologic information is available, including reservoir dimension, a rough range of permeability, the porosity-permeability correlation from AGD (Fig 1(b)), and relative permeability curves of the reservoir rock. We hypothesize that a self-affine fractal model for heterogeneity representation can provide a parameter space that is rich enough to (1) accommodate the target observation data and to (2) yield adequate forecast for reservoir responses. The above hypotheses remain to be tested in the rest of the paper. In other words, *the validity or adequacy of the chosen underlying model space will be evaluated against the utility objectives of the workflow, instead of against the ground-truth reservoir model, which are generally poorly constrained in practice.*

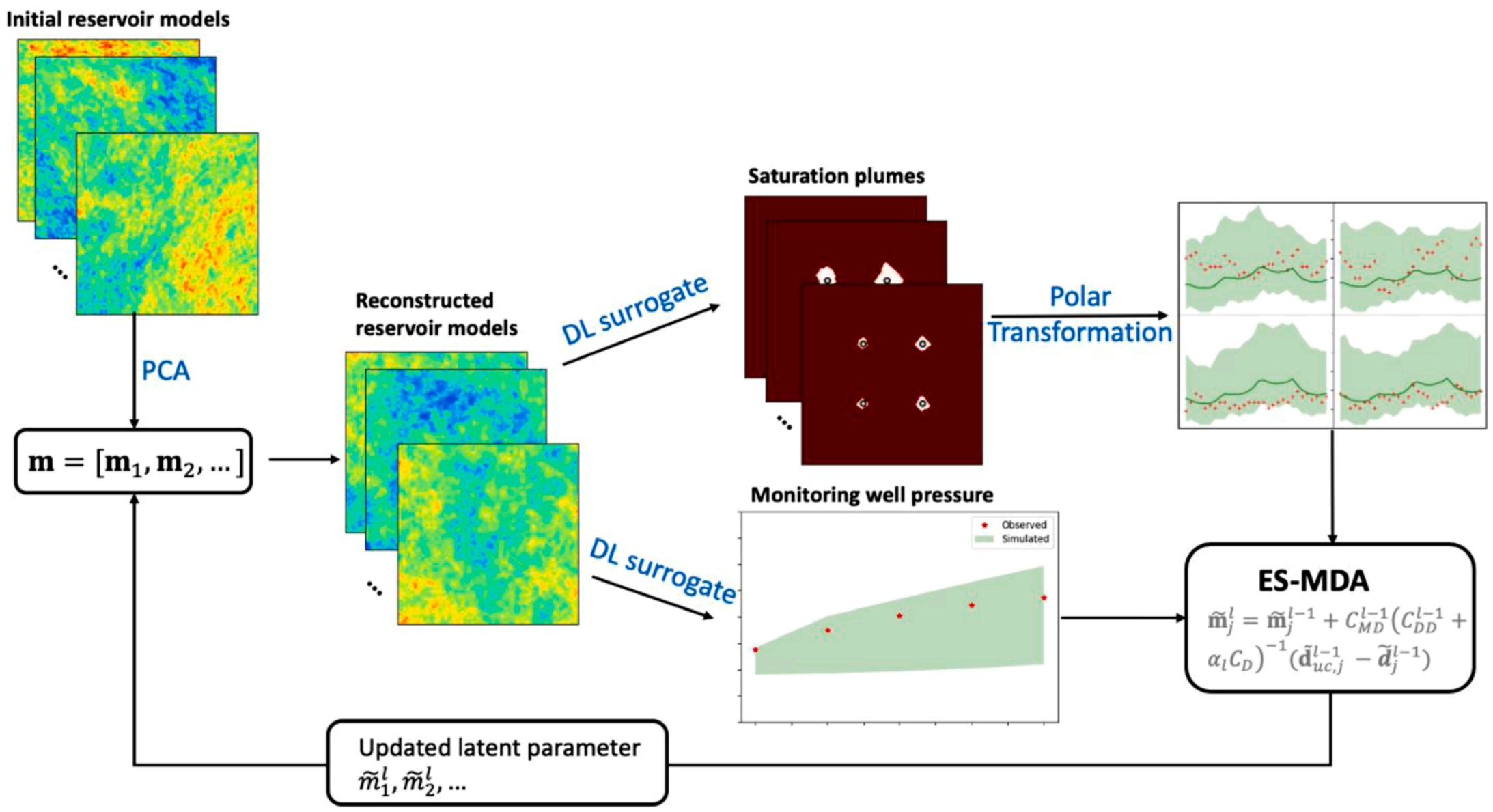

**Fig. 2.** Overview of the data assimilation workflow.

A fractal distribution honors scale-invariant features common for geologic systems (Browaeys and Fomel, 2009; Turcotte and Brown, 1993). Following the method in Sherman et al. (2014) to generate a random field following a fractal distribution, a matrix ($G$) of random values following a normal distribution is generated. The matrix is then transformed to $G^*$ through a Fourier transform. In the frequency domain, $G^*$ is multiplied by a spectral filter ($S$) defined as:

$$\mathbf{S} = \left[(a_x\mathbf{K}_x)^2 + (a_y\mathbf{K}_y)^2\right]^{(-0.5-0.5\beta)} \tag{5}$$

where $\mathbf{K}_x$ and $\mathbf{K}_y$ are the wavenumber components, $axe$ and $a_y$ are scaling parameters for generating anisotropic fields, and $\beta$ is the fractal shape factor which controls the smoothness of the generated fields. The fractal field is finally generated by applying an inverse Fourier transform to the results. Fig. 3 shows examples of anisotropic fractal fields ($axe =$ 1.0, $a_y = 0.6$) generated using different values of $\beta$. Algorithm 1 summarizes the detailed steps of creating a random fractal distribution.

Based on the assumption that porosity fields can be derived from permeability fields through known correlations (Fig. 1(b)) for the target depositional environment, we only generate independent permeability fields following a log-normal distribution applying the fractal model. In the following narrative, we will refer to the logarithmic permeability fields as $k' = \log_{10}(k/\mathrm{m}^2)$. Table 2 summarizes the parameter space applied for generating the random $k'$ fields. The mean and standard deviation of $k'$ are determined based on generic knowledge as discussed above. The generated fields are rotated to create randomly oriented anisotropic fields.

If the permeability values and reservoir responses at the 44,521 (211 $\times$ 211) individual grid cells are directly used as the parameters and data, respectively, the computational cost of the ES-MDA workflow becomes prohibitively high. In the following subsections, we introduce strategies to reduce the dimensionality of the parameter and data spaces.

### 3.4. Parameter space dimension reduction

Instead of directly using the fractal logarithmic permeability fields ($N_k = 44{,}521$ dimensions), we apply a principal component analysis (PCA) approach to generate low-dimensional representations of the fields. Following the implementation details in Liu et al. (2019), we first generate $N_r$ realizations of the fractal logarithmic permeability fields ($k'$) and assemble the vectorized $\mathbf{k}'$ into a matrix $\mathbf{X} \in \mathbb{R}^{N_k \times N_r}$,

$$\mathbf{X} = \frac{1}{\sqrt{N_r - 1}}\left[\mathbf{k}'_1 - \overline{\mathbf{k}}',\ \mathbf{k}'_2 - \overline{\mathbf{k}},\ \ldots,\ \mathbf{k}'_{N_r} - \overline{\mathbf{k}}'\right] \tag{6}$$

$\overline{\mathbf{k}}'$ is the mean of the $N_r$ realizations of $\mathbf{k}'$. A singular value decomposition is performed on $X$ with $\mathbf{X} = \mathbf{U\Sigma V^T}$, where $\mathbf{U} \in \mathbb{R}^{N_k \times N_r}$ and $\mathbf{V} \in \mathbb{R}^{N_r \times N_r}$ are the left and right singular matrices, and $\boldsymbol{\Sigma} \in \mathbb{R}^{N_r \times N_r}$ is a diagonal matrix with its entries known as singular values ($\Sigma_{ii}$) of $\boldsymbol{\Sigma}$. To reduce the original parameter space to an $n$-dimensional space, we select the first $n$ largest components in $\Sigma_{ii}$ and the corresponding columns in $\mathbf{U}$. A new random field (referred to as a PCA realization) can be generated for any given low-dimensional vector $\xi_n \in \mathbb{R}^{n\times 1}$ sampled from $N(0, \mathbf{I}_n)$.:

$$\mathbf{k}'_{\mathrm{PCA}} = \overline{\mathbf{k}}' + \mathbf{U}_n\boldsymbol{\Sigma}_{nn}\boldsymbol{\xi}_\mathbf{n} \tag{7}$$

Eq. (7) defines a PCA model for generating new realizations with features similar to those in the original $N_r$ realizations. The choice of $n$ can be determined by an energy criterion $\sum_{i=1}^{n}\Sigma_{ii} / \sum_{i=1}^{N_r}\Sigma_{ii} \geq T$, where $T$ represents a tolerance value for "energy" preservation. A larger $T$ value results in more faithful representations but also a higher latent space dimension ($n$) and an increased computational expense.

**Table 2**
Parameters for generating fractal logarithmic permeability fields ($k'$). $U$ denotes the uniform distribution.

| Dimension | Mean | Standard deviation | Fractal shape factor ($\beta$) | Scaling parameters | Orientation Angle |
|---|---|---|---|---|---|
| 211 $\times$ 211 | $U(-14, -13)$ | 0.5 | $u(-0.5, 0)$ | $axe = 1.0$ $a_y = 0.6$ | $U(0, 180^\circ)$ |

### 3.5. Data space dimension reduction

Similar to the parameter space dimension reduction, we need to generate low-dimensional representations of $CO_2$ plumes. For binary fields like $CO_2$ plumes, the spatial correlations cannot be characterized by two-point statistics (i.e., by a covariance matrix), so PCA is inapplicable. We propose a polar transformation method to address this challenge. In this method, we transform each plume from the cartesian domain to a polar domain with the well location as the origin. The plume shape can be represented by its radius at angles ($\theta$) with equal intervals in the polar domain. In Fig. 4, we present an example where we transform the original plume of four wells in Fig. 4(a) to a radius vector ($\lambda$) containing 120 variables with each plume represented by 30 values. As shown in Fig. 4(b), the 30 radius variables are equally sampled along 360°. We can also conveniently reconstruct the plume from $\lambda$ as shown in Fig. 4(c). This method is only applicable to "star-shaped" (Stanek, 1977) plumes that do not overlap with each other. Although 2D $CO_2$ plume projections usually fall in this category, some exceptions do exist (Cavanagh, 2013; Hermanrud et al., 2009) and would require different treatments.

### 3.6. The data assimilation workflow

We develop a data assimilation workflow using single-layer 3D reservoir models to match observation data generated from a multi-layer (i.e., vertically discretized) 3D reservoir simulation. A major challenge associated with this simplification is the need to bridge between $CO_2$ plume predictions of single-layer and multi-layer reservoir models. As shown in Fig. 1(c), $CO_2$ plumes have highly variable lateral extents at different reservoir layers. In contrast, a single-layer reservoir model inherently implies a plume with the same extent across the entire depth of the reservoir. We re-emphasize that currently available monitoring technologies like repeat 3D seismic cannot effectively resolve complex vertical structures of $CO_2$ plumes as shown in Fig. 1(c). The variation only became visually apparent in Fig. 1(c) when the vertical dimension was exaggerated by 100 times. Therefore, to simulate $CO_2$ plume size consistent with the observed $CO_2$ plume projection in Fig. 1(e), we introduce an additional parameter ($\chi$), the "porosity scale factor", to the porosity ($\phi$) – permeability ($k$, in mD) correlation as below:

$$\phi = \left(\frac{k}{0.0009 \text{ mD}}\right)^{0.25} \cdot \frac{\chi}{100} \tag{8}$$

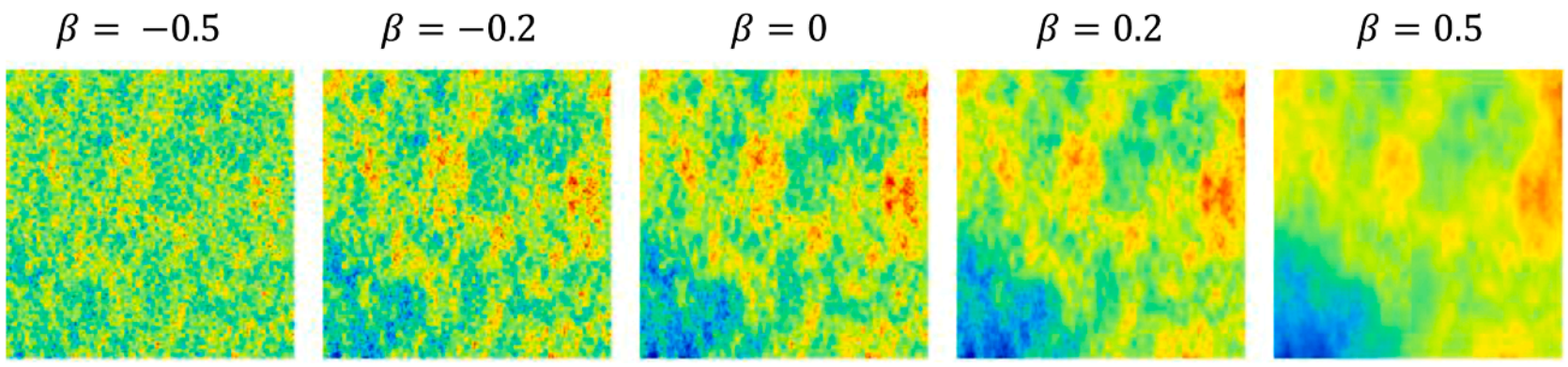

**Fig. 3.** Examples of anisotropic fractal fields generated using different values of fractal shape factor $\beta$.

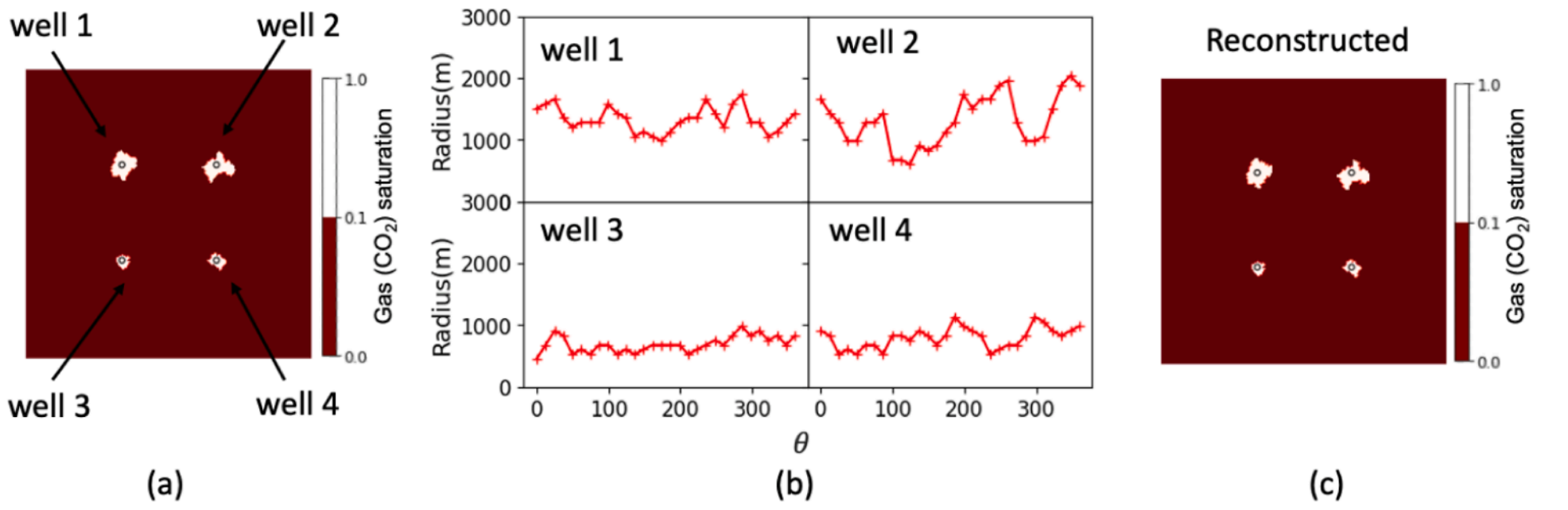

**Fig. 4.** Polar transformation for dimension reduction of $CO_2$ plume. (a) Original $CO_2$ plume map, (b) latent space radius values, and (c) reconstructed $CO_2$ plume map. Note that the black open circles denote well locations.

The physical interpretation of $\chi$ is the ratio of the total pore volume in the multi-layer $CO_2$ plumes to the pore volume in the cylindrical $CO_2$ plumes implied by single-layer models, as illustrated in Fig. 1(d). Since the value of this scaling factor cannot be directly estimated from measurement data, we incorporate $\chi$ as a history matching variable. Notice that the porosity scale factor is only applied to the $CO_2$ plume prediction. For the pressure prediction, we apply the original porosity permeability correlation with $\chi = 1$. It means that we need to perform two forward simulations separately for pressure and saturation estimations.

Algorithm 2 summarizes the data assimilation workflow in detail. The parameters ($m$) in the workflow include $\xi_n$ and $\chi$. The data ($d$) in the workflow include the $CO_2$ plume radius vector ($\lambda$) and the vector ($p$) consisting of five annual measurements of monitoring pressure (vertically averaged reservoir pressure at the monitoring well location). The standard deviations of independent measurement errors for $\lambda$ and $p$ are assumed to be 7.5 m and 0.02 MPa respectively. In this study, the perturbed data are assimilated for $N_a$= 4 steps. The $N_a$ value is chosen based on the recommendation of Emerick and Reynolds (2013) through a trial-and-error process. For each data assimilation step, the forward simulation needs to be run for $2N_e$ times, which is a major computational cost of the workflow. In the next section, we will introduce two DL-based surrogate models to substitute the physical simulators for pressure and $CO_2$ plume predictions.

## 4. Deep learning-based surrogated models

### 4.1. Surrogate model for pressure prediction

For pressure prediction at the monitoring well location, we apply a Wide Residual Network architecture (Zagoruyko and Komodakis, 2016). Compared with the original Residual Network (ResNet) structure in He et al. (2016), Wide ResNet introduces a widening factor $w$ to control the network width. The detailed network structure is given in Fig. 5. Wide ResNet is composed of several basic residual blocks containing batch normalization, ReLU (rectified linear unit) activation, and dropout layers. Batch normalization in the residual block of original ResNet architecture provides a regularization effect. The Wide ResNet architecture additionally adds a dropout layer to perturb batch normalization in the next residual block to prevent overfitting.

The pressure prediction network (WRN-1) requires two input channels: a normalized logarithmic permeability map ($k'$) and a channel indicating the time of prediction ($t$). The normalized permeability maps are downscaled from the original size of 211 × 211 to 64 × 64 using bilinear interpolation. Image resizing is an important preprocessing step in convolutional neural networks that allows various images to meet the requirement of model architectures and to reduce training costs. The final image size is determined by balancing between training cost and model accuracy. To convert the time information to meet Wide ResNet's requirement for input format, we construct a map of the same size as the permeability map but set all the pixel values to a normalized time vale (normalized by the maximum time in the training dataset). The output of the network is a scalar representing the normalized pressure value at the specified time. The normalization is conducted based on the minimum and maximum values in the training and validation datasets. A dropout layer is added into the basic residual block after ReLU activation to perturb batch normalization in the next basic residual block, limiting the potential for overfitting (Zagoruyko and Komodakis, 2016). For this network, we selected a widening factor of $w = 1.0$ and a dropout rate of 0.4. The final constructed network has a total parameter size of 174,337.

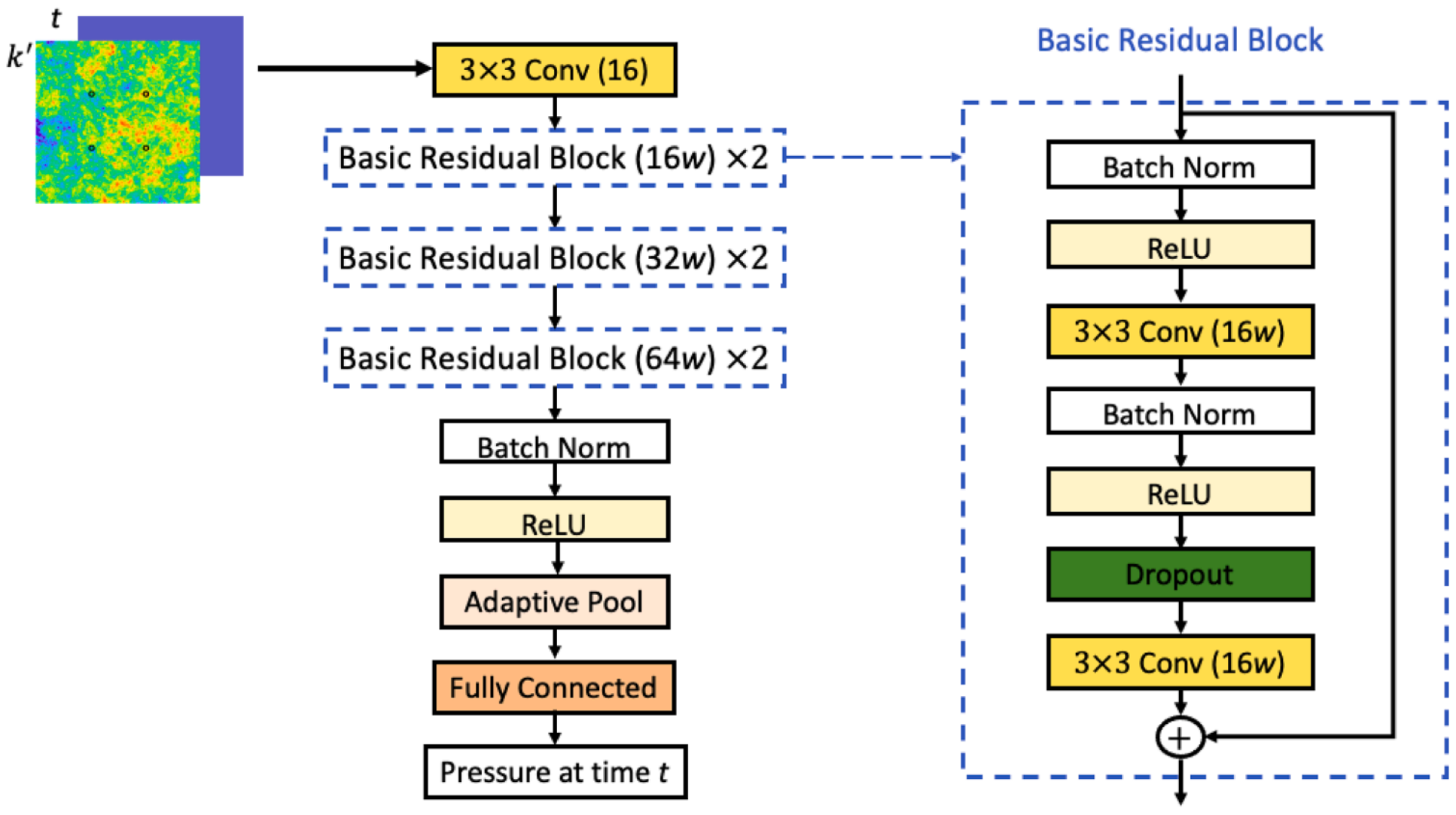

**Fig. 5.** (a) Detailed Wide Residual Network structure for dynamic pressure prediction at the monitoring well location. The input channels include a normalized logarithmic permeability map ($k'$) and a constant map indicating time ($t$). The notation "3 × 3 Conv (16)" denotes a convolutional layer with a kernel size of 3 × 3 and 16 channels. The widening factor ($w$) controls the number of channels in basic residual blocks.

### 4.2. Surrogate model for $CO_2$ plume prediction

For a multi-well injection scenario, wells are subjected to different injection rates determined by the injectivities of individual wells. Consequently, depending on the geologic realization, each of the four injection wells could have different injection rates of $CO_2$. We propose a two-step strategy, each step employing a different neural network, to build a DL-based surrogate model for $CO_2$ plume prediction (Fig. 6).

The first step predicts injection volume allocations among the wells, quantified by four injection rate allocation ratios ($\eta_i$), which are the ratios of injected quantities into individual wells at the time of evaluation to the total injection quantity for the well group. We use a second Wide ResNet (WRN-2) for this step, which takes the normalized permeability maps as the main input. We selected the widening factor and the dropout rate to be 2.0 and 0.4 respectively, and we modified the activation function of the output layer to be "softmax" to reflect the constraint that the sum of the four output values is equal to unity. WRN-2 has a total parameter size of 690,612.

The second step predicts $CO_2$ plume shape and sizes around the four wells using a R-U-Net (Tang et al., 2020). R-U-Net denotes "residual U-Net", where basic residual blocks in ResNet (He et al., 2016) are added into a U-Net architecture (Long et al., 2015). The U-Net architecture is featured as a fully convolutional networks (without fully connected layers) and therefore can handle large input maps. The R-U-Net architecture implemented in this paper is given in Fig. 7. There are two input channels for this R-U-Net. The first is a normalized logarithmic permeability map ($k'$), which is cropped from 211 × 211 to 200 × 200 so that the margins of the permeability maps do not influence the $CO_2$ plume shapes. The other input channel is a map of the same size as the permeability map. We give scalar values $\eta_i t/\chi$ to the pixels at the well locations, and other pixel values are set to zero. The given values indicate the plume size for each well at different times ($t$) and porosity scale factors ($\chi$). The encoding network extracts feature maps from input maps through several convolutional layers. The last feature maps are passed to six residual blocks and then fed into the decoding network. The feature maps are upsampled through de-convolutional layers in the decoding network to obtain a final $CO_2$ plume map. The concatenation layers (gray arrows in Fig. 7) are an important feature of the U-Net structure. These layers copy feature maps in the encoding net and concatenate them with the up-sampled feature maps in the decoding net. The total parameter size of this network is 607,057. Note that based on our experiments, a direct implementation of R-U-Net with permeability maps alone as input cannot give reasonable predictions of $CO_2$ plumes for multi-well injection.

### 4.3. Data sets and training procedure

An open source multiphysics simulator GEOSX (http://www.geosx.org/) is applied to generate training datasets for the deep learning-based surrogate models. Table 3 summarizes the key reservoir properties used in GEOSX simulation. The reservoir dimension and fluid injection target are the same as the reference model. The relative permeability curve of high permeable sandstone facies in the reference model is adopted. Instead of using a full EOS model (Peng-Robinson EOS), we apply a $CO_2$-brine model in GEOSX. This simplified EOS model considers two components ($CO_2$ and $H_2O$) in two phases (brine phase and $CO_2$ phase). The flash calculation is based on the model of Duan and Sun (2003).

We generate 1080 initial fractal permeability fields using the parameters indicated in Table 2. A PCA model is built for generating new permeability fields honoring the features of the initial fractal fields. The 996 principal components are selected based on a tolerance value $T = 0.99$. We generate 10,800 permeability fields in total for training and testing the surrogate models. For each PCA-generated permeability field, we run two forward simulations with GEOSX and output the simulation results annually for ten years. One forward simulation is for pressure prediction with porosity scale factor $\chi = 1.0$, and the other simulation is for $CO_2$ plume prediction with $\chi$ uniformly sampled between (0.05, 0.45). The training process is to minimize the $L_a$ norm of the difference between GEOSX model predictions ($\mathbf{y}_i$) and neural network predictions ($\widehat{\mathbf{y}}_i$),

$$L_a = \frac{1}{n_m}\frac{1}{n_t}\sum_{i=1}^{n_m}\sum_{t=1}^{n_t}\| \widehat{\mathbf{y}}_i^t - \mathbf{y}_i^t \|_a^a \tag{9}$$

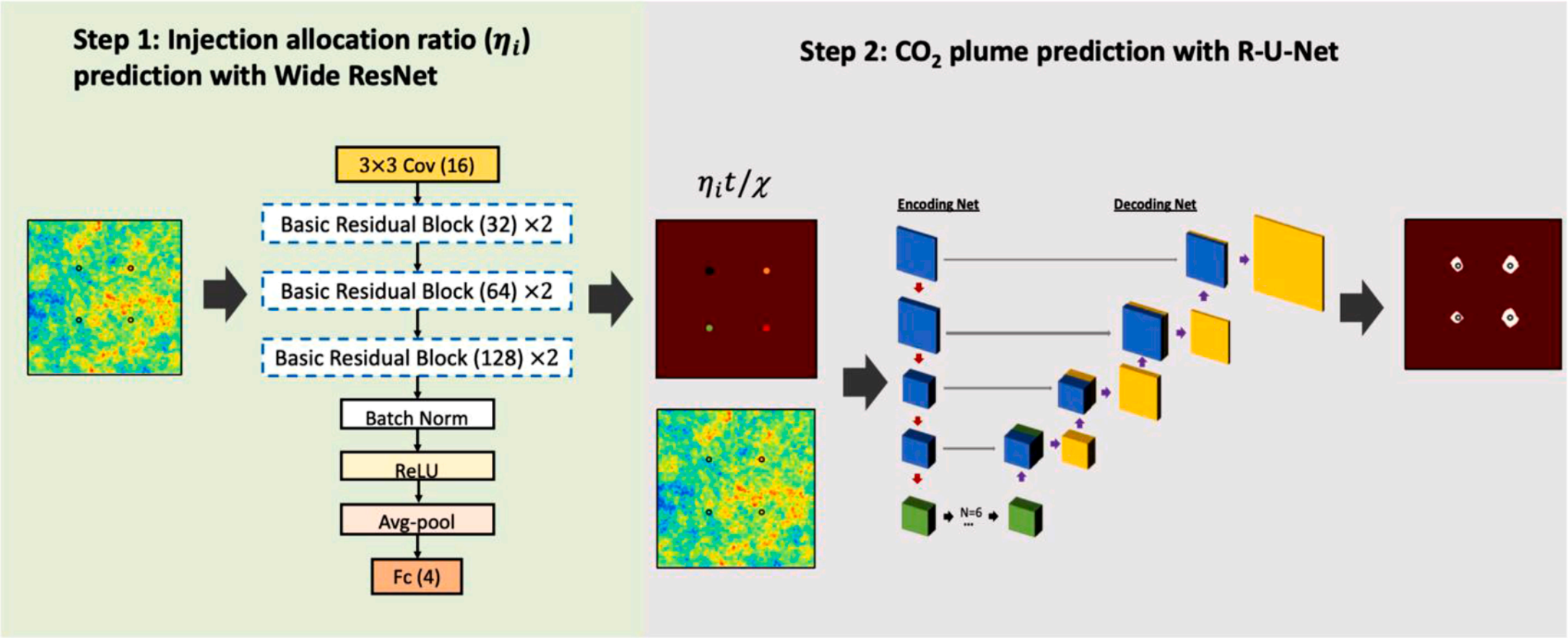

**Fig. 6.** Illustration of the two-step approach for building the DL-based surrogate models for $CO_2$ plume prediction.

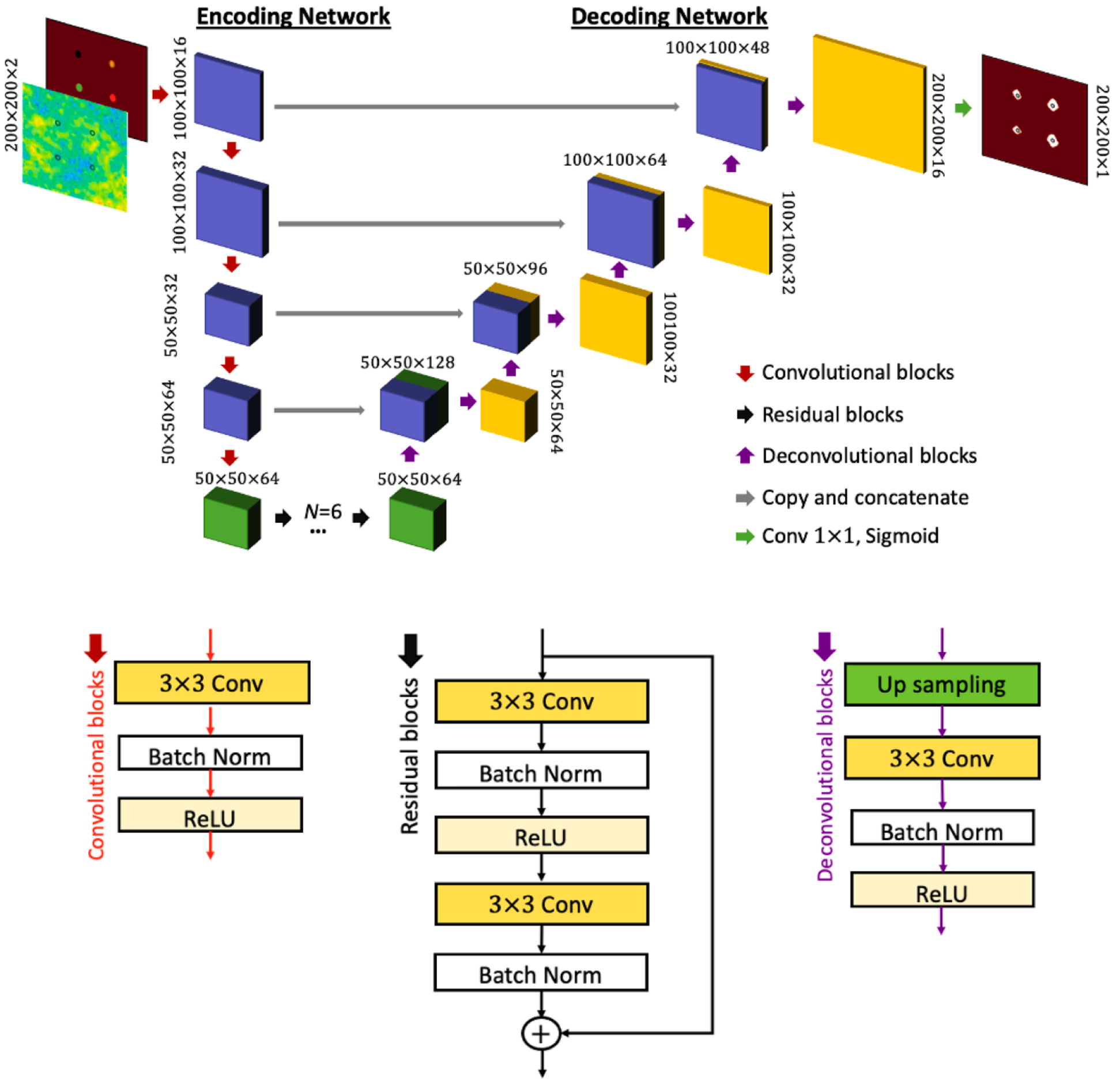

**Fig. 7.** Illustration of the R-U-Net architecture in the second step of the surrogate model for $CO_2$ plume prediction. The input channels include a normalized logarithmic permeability map and a map with pixel values at the well locations being $\eta_i t/\chi$ and other pixel values being zero. $N = 6$ indicates 6 residual blocks.

**Table 3**
Reservoir properties used in GEOSX simulation.

| Parameters | Value |
|---|---|
| Reservoir dimension | 32,156.4 $m$ × 32,156.4 $m$ × 91.4m |
| Grid division | 211 × 211 × 1 |
| Field injection target | 2 million metric tons/per year |
| Compositional fluid model | $CO_2$-brine model (Duan and Sun, 2003) |
| Relative permeability | High permeable sandstone (Bennion and Bachu, 2007) |

where $n_m$ is the number of training permeability maps (between 1000 and 10,000) and $n_t$ is the number of time steps in the training data sets ($n_t = 10$ for WRN-1 and R-U-Net; $n_t = 1$ for WRN-2). $\widehat{\mathbf{y}}_i$ and $\mathbf{y}_i$ are predictions of neural networks and GEOSX, respectively. For WRN-1, $\mathbf{y}$ is the dynamic pressure at the monitoring well location. For WRN-2, $\mathbf{y}$ is the calculated injection allocation ratio. For R-U-Net, $\mathbf{y}$ is the vectorized saturation map. We found that better prediction results can be achieved by using the $L_2$ norm for training WRN-1 and the $L_1$ norm for training WRN-2 and R-U-Net.

We applied the ADAM optimization algorithm (Kingma and Ba, 2014) with an initial learning rate of 0.001 and a batch size of 64 to train the networks. Both WRNs converged within 300 training epochs and the R-U-Net converged within 100 training epochs. The networks are trained on a Nvidia Tesla P100 GPU for 30–260 min, 4–33 min, and 63–600 min for WRN-1, WRN-2, and R-U-Net, respectively. We use three metrics to evaluate network accuracy: the root mean square error (RMSE), the coefficient of determination ($R^2$), and the structural similarity index measure (SSIM):

$$\mathrm{RMSE} = \sqrt{\frac{1}{n_s}\frac{1}{n_t}\sum_{i=1}^{n_s}\sum_{t=1}^{n_t}\| \widehat{\mathbf{y}}_i^t - \mathbf{y}_i^t \|_2^2} \tag{10}$$

$$R^2 = 1 - \frac{\sum_{i=1}^{n_s}\sum_{t=1}^{n_t}\| \widehat{\mathbf{y}}_i^t - \mathbf{y}_i^t \|_2^2}{\sum_{i=1}^{n_s}\sum_{t=1}^{n_t}\| \widehat{\mathbf{y}}_i^t - \overline{\mathbf{y}}_i^t \|_2^2} \tag{11}$$

where $n_s = 400$ is the number of testing permeability maps. SSIM is applied to quantify the similarity between two images and is calculated as (Wang et al., 2004):

$$\mathrm{SSIM}(\mathbf{u},\ \mathbf{v}) = \frac{1}{M}\sum_{m=1}^{M}\frac{\left(2\mu_{u,m}\mu_{v,m} + K_1^2\right)\left(2\sigma_{uv,m} + K_2^2\right)}{\left(\mu_{u,m}^2 + \mu_{v,m}^2 + K_1^2\right)\left(\sigma_{u,m}^2 + \sigma_{v,m}^2 + K_2^2\right)} \tag{12}$$

SSIM is evaluated based on local windows (with size 11 × 11, denoted by $m$) in the true and predicted images $\mathbf{u}$ and $\mathbf{v}$. $M$ is the number of local windows of the image. $\mu$ and $\sigma$ denotes the mean and standard deviation of the vector formed by local windows, respectively. $K_1 = 0.01$ and $K_2 = 0.03$ are two small values introduced to avoid denominator failure. A mean SSIM (MSSIM) is calculated to evaluate the overall accuracy of testing data set.

$$\mathrm{MSSIM} = \sum_{i=1}^{n_s}\mathrm{SSIM}(\mathbf{u_i},\ \mathbf{v_i}) \tag{12}$$

A lower RMSE value, and an $R^2$ value and a MSSIM value approaching 1.0 indicate better model accuracy.

### 4.4. Evaluation of surrogate model performance

WRS-1 is applied as the surrogate model for pressure prediction, and a combination of WRS-2 and R-U-Net are combined as the surrogate model for $CO_2$ plume prediction as shown in Fig. 6. We first test the accuracy of surrogate models trained with different training sample sizes, $n_m$ =1000, 2000, 5000 and 10,000. Fig. 8(a) shows the RMSE and $R^2$ metrics for pressure prediction, and Fig. 8(b) shows the RMSE and MSSIM metrics for saturation prediction. As $n_{\mathrm{m}}$ increases from 1000 to 10,000, pressure model accuracy increases significantly from a RMSE value of 0.046 to 0.027, and the saturation model accuracy only increases from 0.041 to 0.035. The results also indicate that using 1000 training maps can achieve comparable accuracy (MSSIM= 0.985) as using 10,000 training maps (MSSIM= 0.988) for saturation prediction. Note that the subsequent model evaluations and applications will be based on surrogate modes trained with 10,000 maps although the results based on a smaller training sample size could achieve adequate accuracy.

We further demonstrate model performance in Figs. 9–11, which shows the results for six randomly selected permeability maps, each with a random porosity scaling factor $\chi$. The results include predicted $CO_2$ plumes after five years of injection and the dynamic pressure evaluations. Both saturation and pressure models give satisfactory prediction results with SSIM $> 0.95$ and $R^2 > 0.98$. Note that the pressure predictions do not take $\chi$ as inputs. As $\chi$ increases from 0.05 to 0.45, the $CO_2$ plume size decreases significantly from a large extent to several pixels around wellbores.

We test the generalizability of the surrogate models on permeability fields outside the testing data set with different feature richness. The porosity scale factor is 0.2 for the following testing cases. As shown in Fig. 10, we apply different PCA tolerance values $T = 1.00$, 0.99, 0.90, 0.80, 0.70 and 0.60 for generating input permeability fields. The corresponding latent space dimensions are 1079, 927, 404, 188, 79 and 28 respectively. As $T$ decreases, the generated permeability fields have a decreasing feature richness and an increasing smoothness. Although the surrogate models were trained based on data generated from $T = 0.99$, the SIMM values of saturation predictions at the 5th year are all above 0.97, and the $R^2$ values for all pressure predictions are equal to or above 0.97. Fig. 9 also indicates that as the feature richness of the permeability map increases, the saturation plumes tend to have rougher perimeter profiles and thereby more "complex" shapes. This finding motivates us to use a high PCA tolerance value for data assimilation in the next section.

Fig. 11 presents the dynamic saturation predictions for the permeability map ($T = 0.99$) shown in Fig. 10. For injection duration of 2.5, 5, 7.5, 10, 15 and 20 years, the SSIM values are 0.981, 0.986, 0.978, 0.973, 0.966 and 0.952 respectively. Note that the surrogate model is trained based on simulation results recorded annually for 10 years. The results illustrate that the surrogate model performs well at interpolation within the training dataset and can extrapolate reasonably well into the future.

## 5. Data assimilation and forecasting results and discussion

### 5.1. ES-MDA inversion and forecasting results

We follow Algorithm 2 in Section 3.4 to perform data assimilation and forecasting. We use an ensemble size $N_e = 10{,}000$ to evaluate the uncertainties of this high-dimensional inverse problem. We notice that an ensemble size smaller than 5000 tends to lead to ensemble collapse or failure of the algorithm. To generate fields with enough richness to match the observed saturation plumes, we apply a PCA tolerance $T = 0.99$ (number of principal components $n = 996$). The parameter space dimension is 997 (with porosity scale factor), and the data space dimension is 125 (120 variables to capture the plume shapes and five pressure values) in this problem.

Fig. 12 presents eight randomly selected realizations from the 10,000 prior and posterior permeability realizations, as well as the mean and standard deviation maps of the prior and posterior ensembles calculated for the entire ensemble. As explained in Section 3.3, we do not have a ground-truth permeability map against which to compare the results. The performance of the workflow is evaluated by comparing the $CO_2$ plume and pressure predictions from history-matched models to those generated from the original 3D reservoir simulation results. The posterior permeability fields have higher permeability around injection wells 1 and 2 locations than that around injection wells 3 and 4 locations. This was apparently driven by the larger observed plume sizes around injection wells 1 and 2 than around injection wells 3 and 4 as shown in Fig. 3. The uncertainty of permeability estimation around the wellbore locations is much lower than that of the far field regions presumably because $CO_2$ plume shapes within the first 10 years of $CO_2$ injection are primarily affected by properties in the near-well regions.

Fig. 13(a) shows the box plot of porosity scale factor ($\chi$) at each ES-MDA iteration. The red line indicates an estimated value of $\chi$ ($\approx 0.13$) based on the 3D simulated plumes of the reference model and the 2D observed plumes. The plot illustrates that the data assimilation workflow starts to converge after three iterations. Fig. 13(b) shows the 90% credible interval of pressure predictions from prior and posterior logarithmic permeability ensembles. The true pressure curve and the observed pressure data (the same as in Fig. 1(e)) are provided in the plot for comparison. The posterior logarithmic permeability ensemble matches the observation data with better accuracy and significantly reduces the uncertainty of dynamic reservoir pressure predictions at the monitoring location between five to ten years.

Fig. 14 presents $CO_2$ plume predictions after five and ten years of injection based on the prior and posterior $k^{'}$ fields ensembles. The latent space radius vectors are used for quantitative comparison between the "true" $CO_2$ plume and the predicted $CO_2$ plume. As expected, the $CO_2$ plume predicted from the posterior log permeability ensemble can accurately match the observed $CO_2$ plumes both quantitively and qualitatively (Fig. 14(a)). Fig. 14(b) illustrates that the posterior log permeability fields can provide reasonable forecasts of $CO_2$ plume migration for future injection periods.

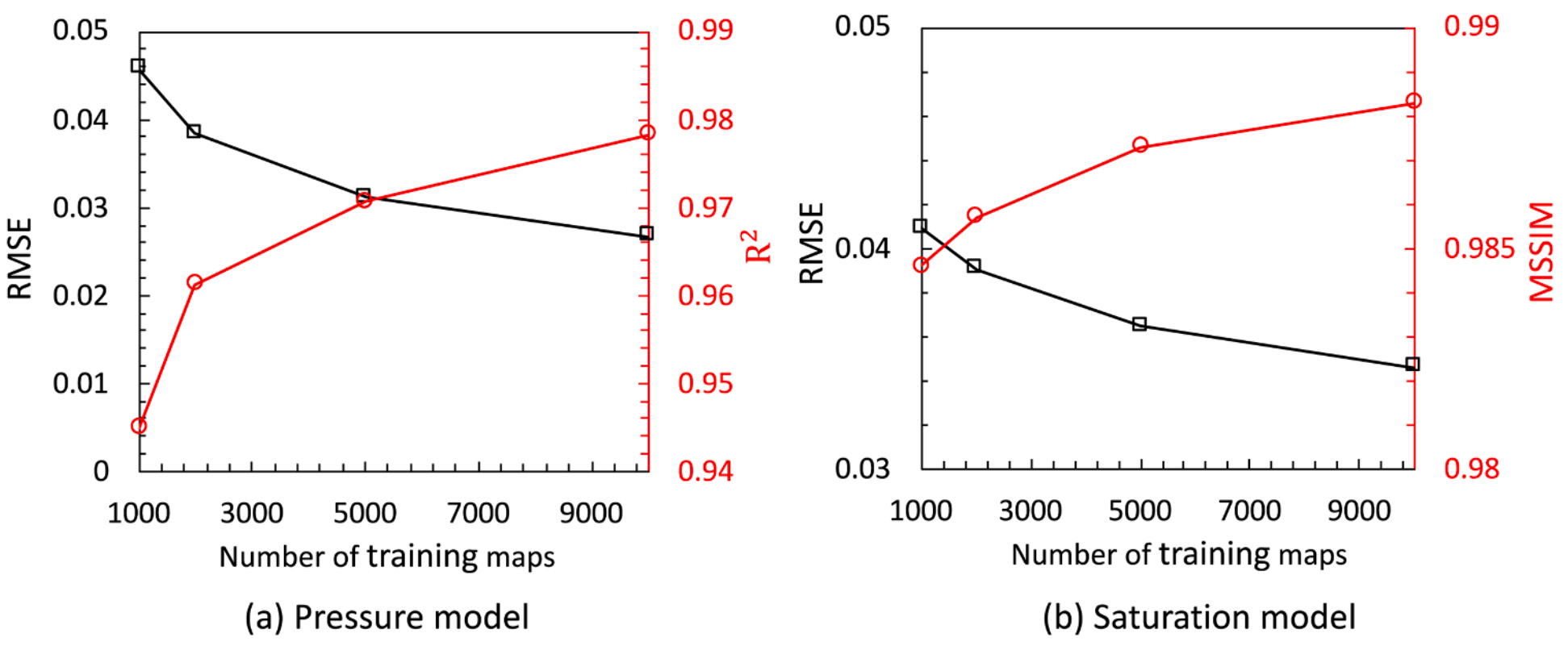

**Fig. 8.** Performance metrics of (a) pressure surrogate model and (b) saturation surrogate model.

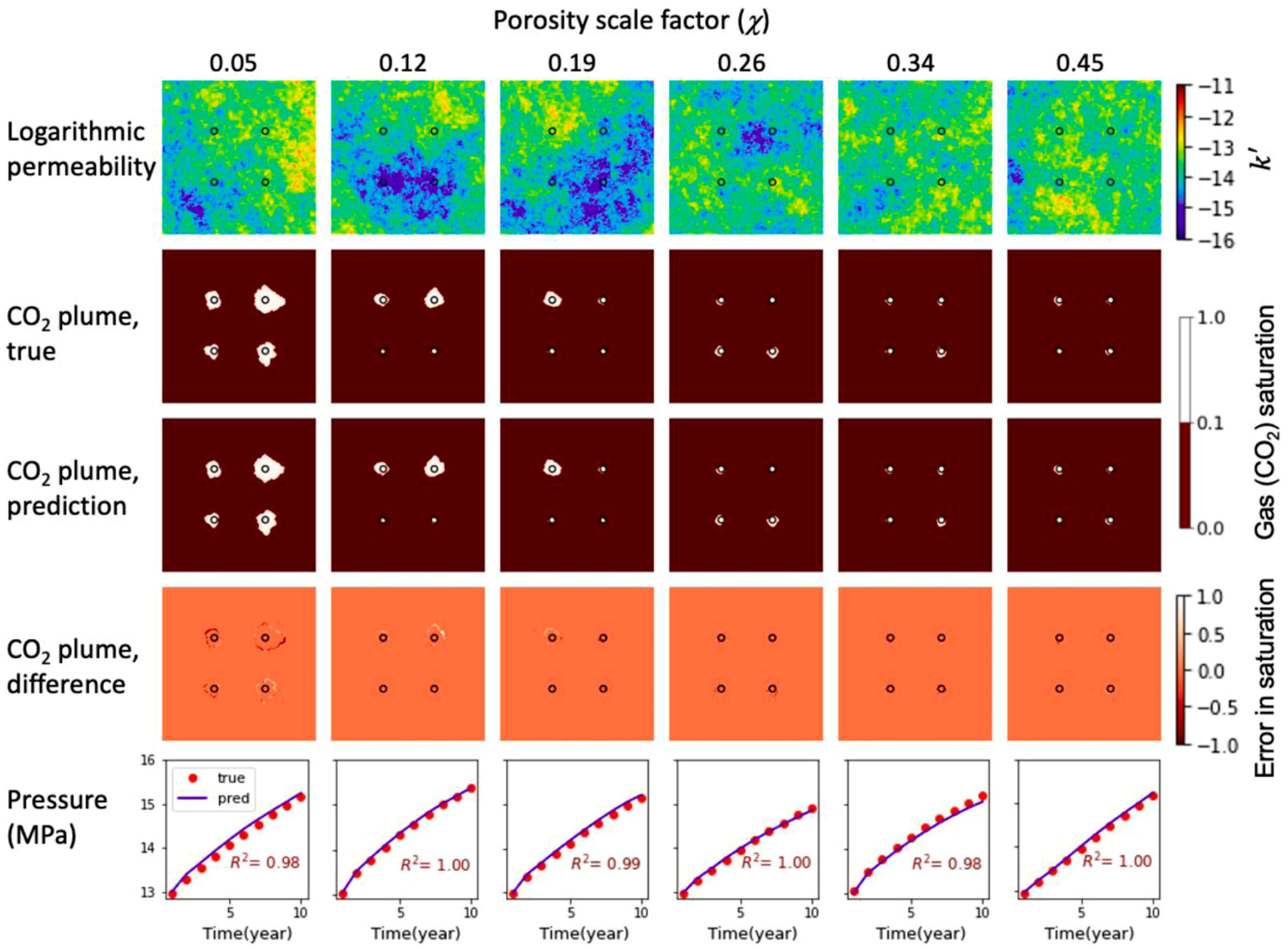

**Fig. 9.** Logarithmic permeability fields and porosity scale factor ($\chi$) sampled from the testing dataset, $CO_2$ plumes predicted by GEOSX (true) and the saturation surrogate model (prediction) after five years of injection, and dynamic pressure at the monitoring well location predicted by GEOSX and the pressure surrogate model.

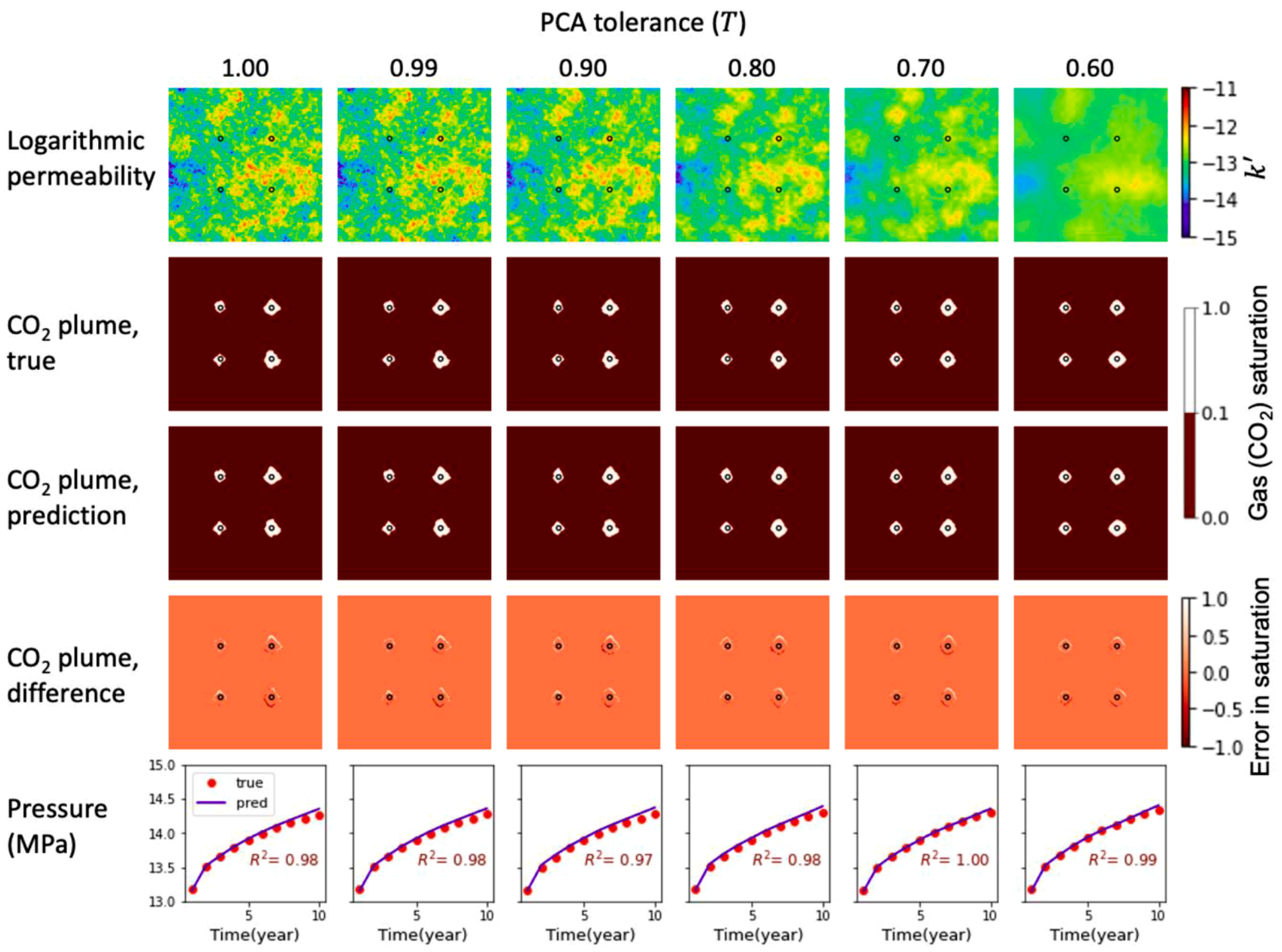

**Fig. 10.** Logarithmic permeability fields generated based on different PCA tolerances ($T$), $CO_2$ plumes predicted by GEOSX (true) and the surrogate model (prediction) after five years of injection, and dynamic pressure at the monitoring well location predicted by GEOSX and the pressure surrogate model.

### 5.2. *Sensitivity analysis on surrogate models with different accuracy*

As a sensitivity analysis, we rerun the data assimilation workflow four times using surrogate models trained based on different numbers of permeability maps ($n_m$= 1000, 2000, 5000 and 10,000). To quantify the data assimilation and prediction performance, we use the RMSE and $R^2$ as given in Eqs. (10) and (11), respectively. In this application, $\hat{\mathbf{y}}_i$ denotes the mean predicted values from the posterior ensemble and $\mathbf{y}_i$ denotes the true values from multi-layer 3D simulation. We evaluate three metrics, each in a vectorial form: pressure for 10 years, as well as radius vectors of $CO_2$ plumes after five and ten years of injection. Fig. 15 presents the evaluation results, which follow the same convention as Fig. 8. Pressure predictions are more sensitive to the number of training permeability maps than the $CO_2$ plume predictions. Among the three metrics, the $CO_2$ plume prediction after ten years of injection is the least sensitive to the accuracy of surrogate models. Similar to Fig.14(b), all the surrogate models (although with different accuracies) are able to predict a reasonable $CO_2$ plume shape after ten years of injection.

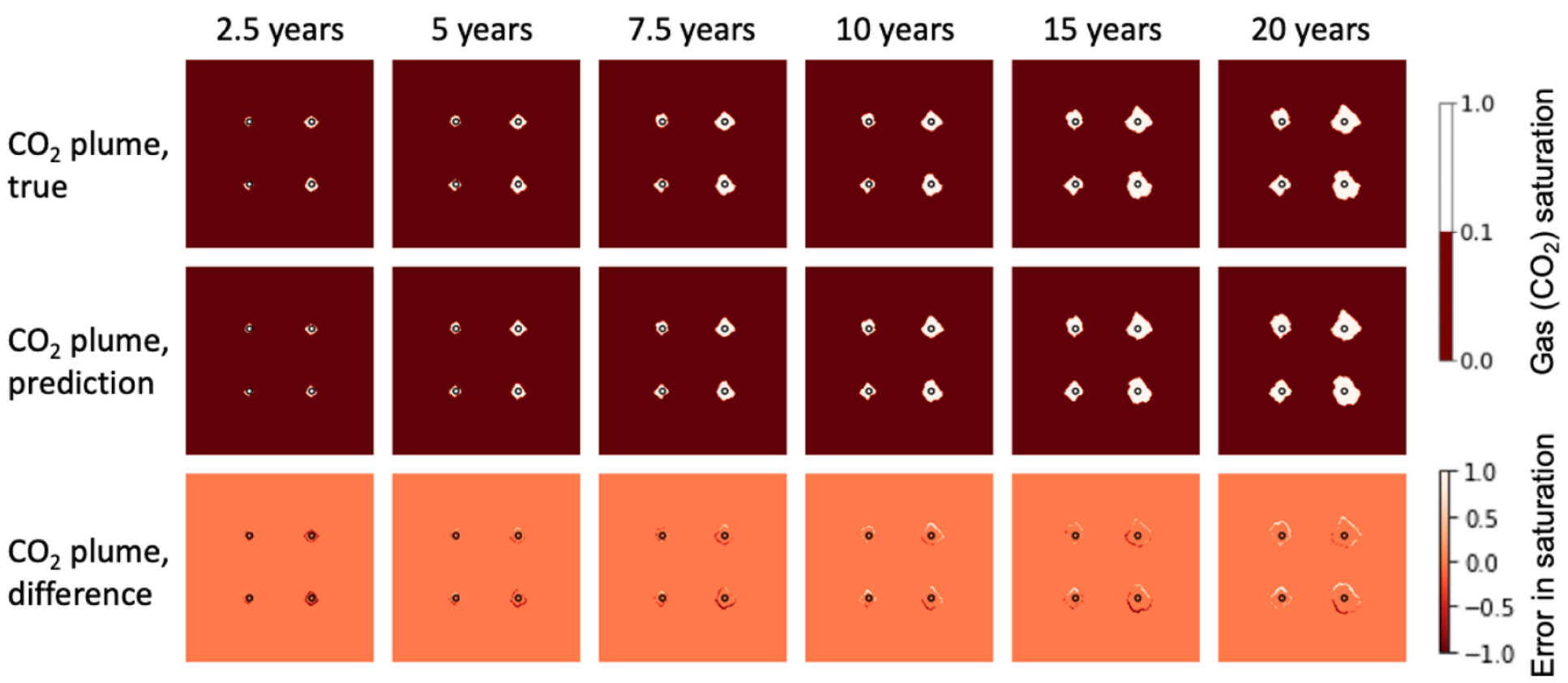

**Fig. 11.** $CO_2$ plumes predicted by GEOSX (true) and the surrogate model (prediction) after different injection durations for the permeability map with $T = 0.99$ in Fig. 9.

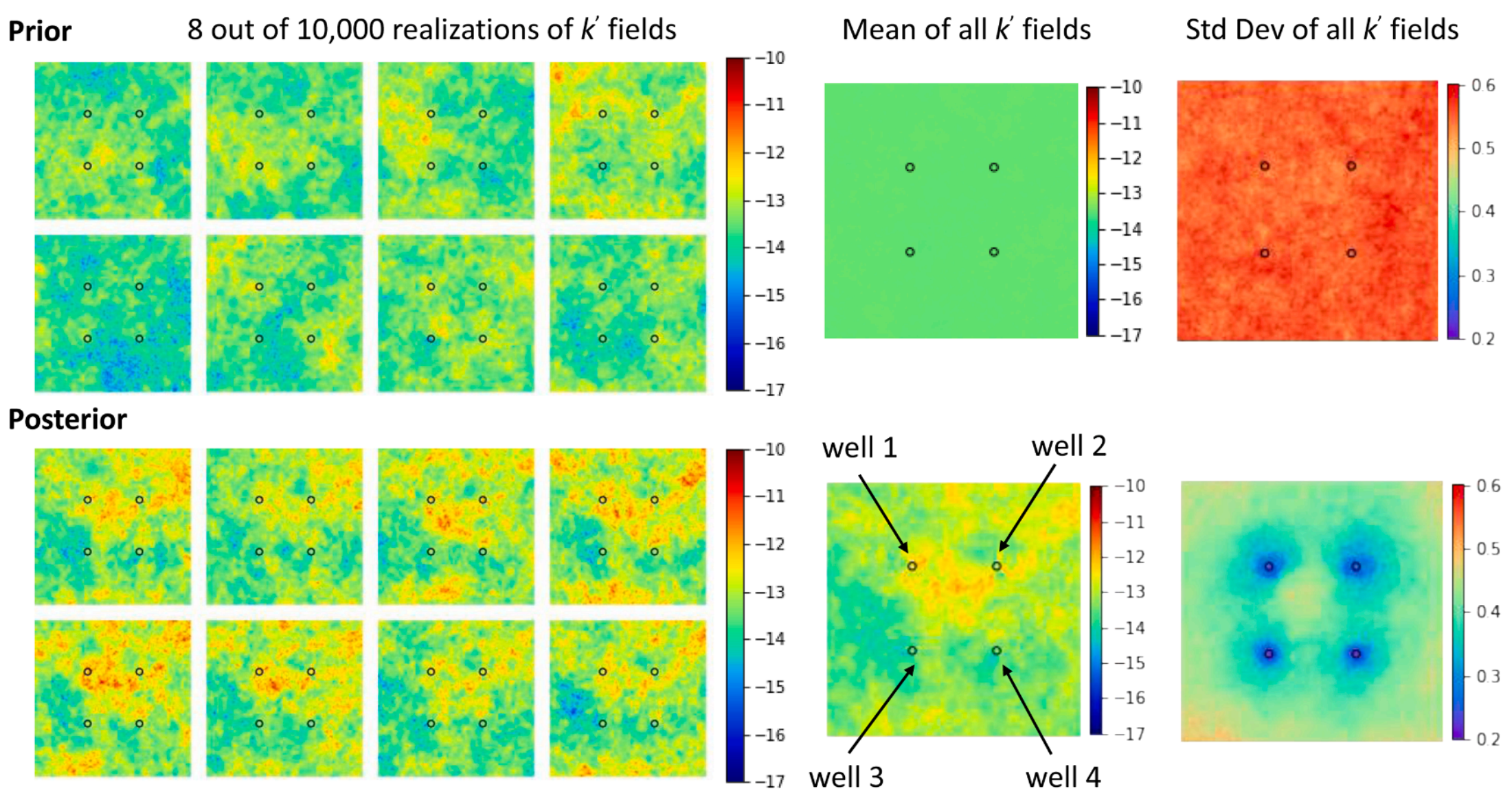

**Fig. 12.** Eight randomly selected (out of 10,000) realizations of prior logarithmic permeability ($k'$) fields and the corresponding posterior $k'$ fields of the data assimilation. The mean and standard deviations of the $k'$ fields are calculated for the entire ensemble.

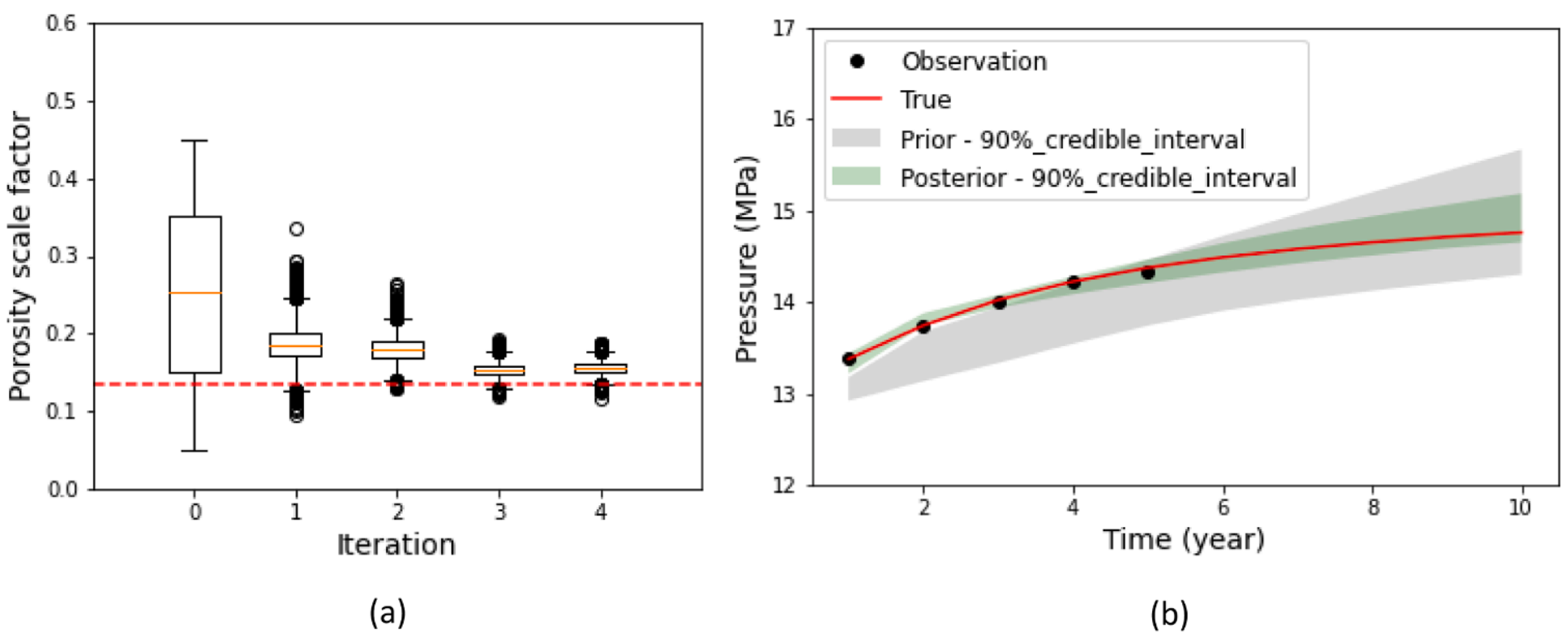

**Fig. 13.** (a) Box plot of porosity scale factor distributions at each ES-MDA iteration. (b) 90% credible intervals of dynamic pressure predictions based on prior and posterior permeability ensembles, along with observed pressure data and true pressure curve based on the reference model.

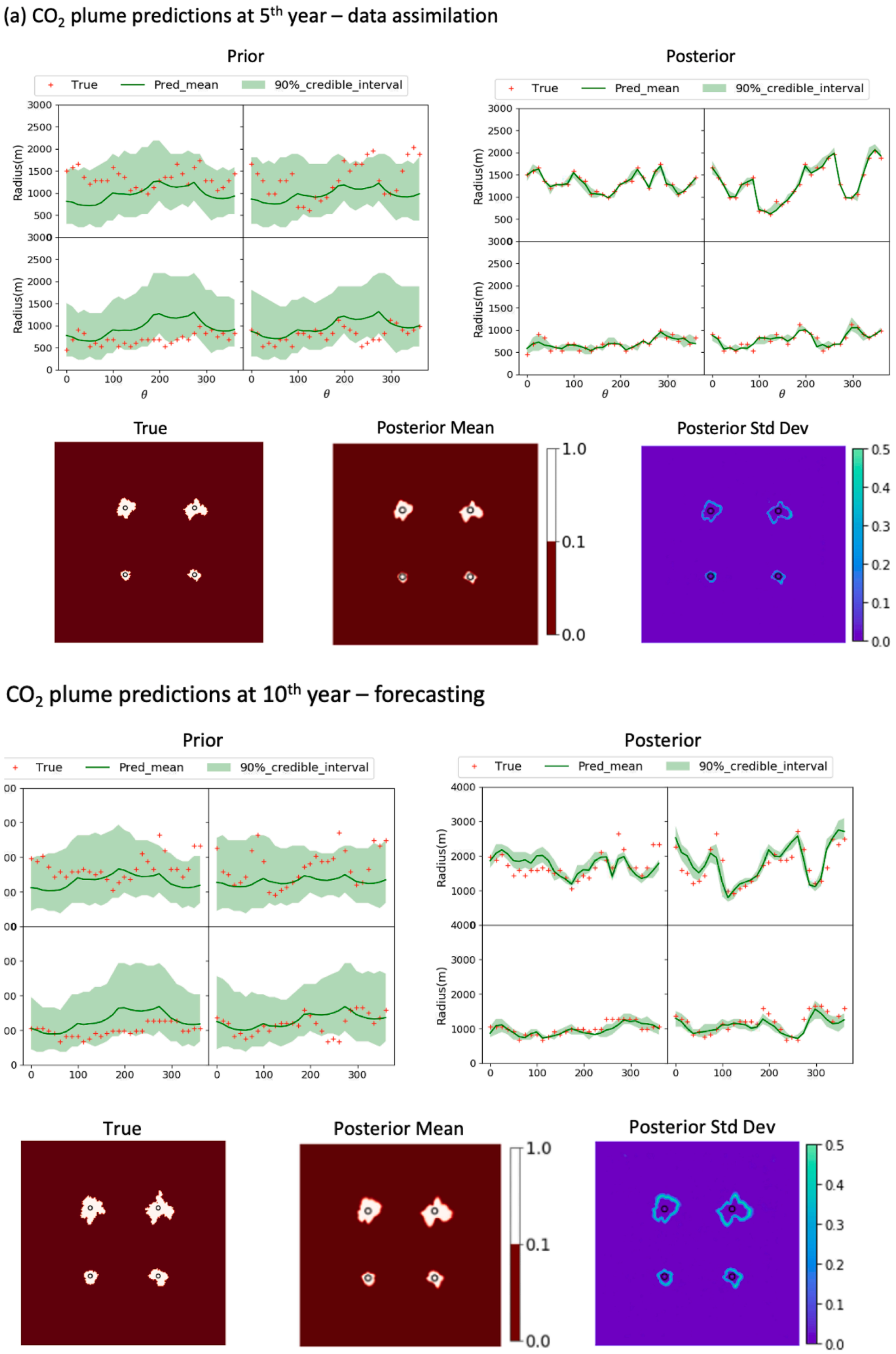

**Fig. 14.** $CO_2$ plume predictions after (a) five years and (b) ten years of injection. Prior and posterior predictions are compared with true $CO_2$ plume through radius obtained through polar transformation.

### 5.3. Computational cost of the workflow

In Table 4, we summarize the computational cost of the data assimilation and forecasting workflow. With an ensemble size of 10,000 and four iterations, the workflow requires 80,000 forward simulations in total to evaluate pressure and $CO_2$ plume. If GEOSX is used as the forward simulator, then it will take approximately 13,000 core-hours to run the workflow. In our applications, where forward physical simulators are replaced by the DL-based surrogate models, the workflow runs in approximately 2 h on a 4-core laptop and within one hour on a 12-core workstation. A significant amount of computational cost has been transferred to training surrogate models, which can be completed before

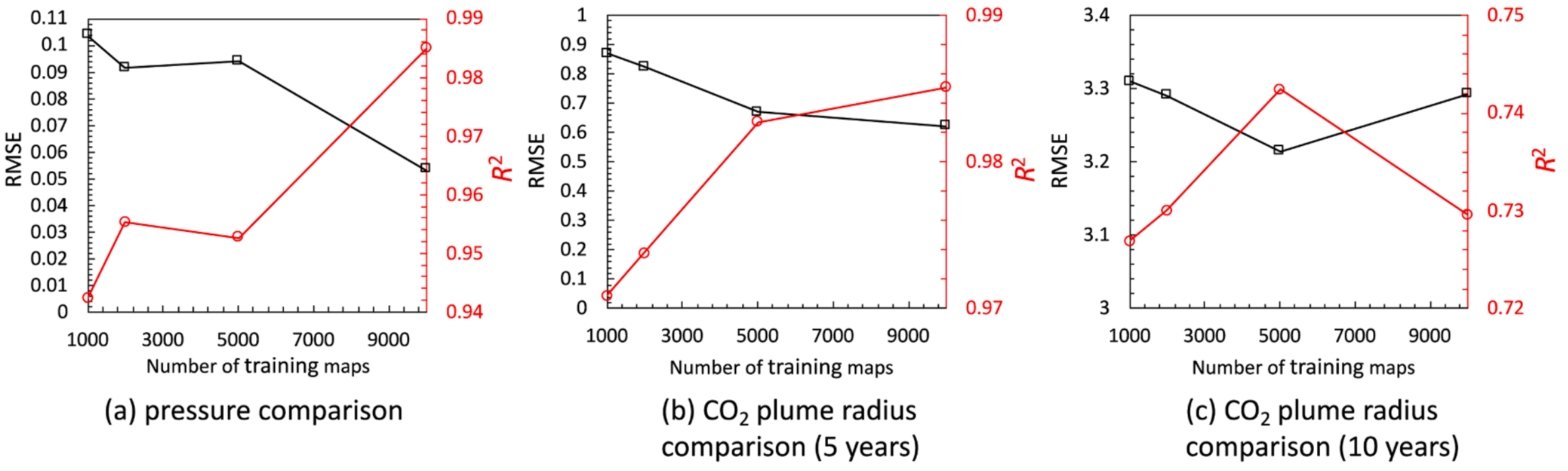

**Fig. 15.** RMSE and $R^2$ metrics of (a) predicted pressure (mean of the entire ensemble) and true pressure and predicted $CO_2$ plume radius (mean of the entire ensemble) and observed $CO_2$ plume radius after (b) 5 years and (c) 10 years of injection.

**Table 4**
Summary of the computational cost of the data assimilation workflow.

| Time | Physical Simulator | DL-Based Surrogate Models |
|---|---|---|
| Single run ($CO_2$ plume) | ~10 core-min[a] | 0.48 core-s[b] |
| Single run (pressure) | ~10 core-min[a] | 0.36 core-s[b] |
| Data assimilation (10,000 ensemble, 4 iteration) | ~13,000 core-hours[a] | <1 machine-hour[c], or ~2 machine-hours[d] |
| Training dataset cost | – | ~3000 core-hours[a] |
| Training | – | 4 GPU-hours / 10.5 GPU-hours[e] (pressure / $CO_2$ plume) |

[a] On an Intel Xeon E5–2695 v4, single-core serial run.
[b] On MacBook Pro, i5–8279 U CPU, single-core run.
[c] On 12-core workstation with Intel Xeon E5–2678 v3, multi-core parallel run.
[d] On MacBook Pro, i5–8279 U CPU, utilizing up to four cores.
[e] On an NVIDIA Tesla P100 GPU.

**Algorithm 1**
Generate a 2D random fractal field in a cartesian coordinate system.

**Inputs**: Dimension of the output field ($n_x$, $n_y$), grid spacing $\Delta x$ (assumed to be equal in the two directions), fractal shape factor $\beta$, and scaling parameters ($a_x$, $a_y$)

1 Calculate wavenumber components $\mathbf{K}_x$ and $\mathbf{K}_y$

$$\mathbf{K}_x = \frac{1}{2\Delta x}\left(\frac{2i}{n_x} - 1\right) \text{ for } i = 1, \ldots, n_x$$
$$\mathbf{K}_y = \frac{1}{2\Delta x}\left(\frac{2j}{n_y} - 1\right) \text{ for } j = 1, \ldots, n_y$$

2 Build the spectral filter $\mathbf{S}$ based on Eq. (5) and shift the zero-frequency component to the center of the spectrum.
3 Generate a random matrix $\mathbf{G} \in \mathbb{R}^{n_x \times n_y}$ and conduct Fast Fourier Transform (FFT) to obtain $\mathbf{G}^*$
4 Conduct inverse Fourier transform to $\mathbf{S} \times \mathbf{G}^*$ and discard imaginary components.
5 Normalize the results from step 4.

the data assimilation process. The training costs include generating the training data set (approximately 3000 core-hours for 10,000 training permeability maps) and training the neural networks (14.5 GPU-hours for 10,000 training permeability maps). The total time taken for completing the workflow (with training time added) is still significantly less than that of applying a physical simulator.

## 6. Discussion

### 6.1. Uncertainty assessment

A major difference of data assimilation for GCS in deep saline aquifers from traditional petroleum reservoir history matching is that we need to deal with more uncertainties. Firstly, we are in general lack of sufficient well data to constrain the geologic description of the storage formations due to the high drilling costs(Chen et al., 2020; Ma et al., 2019). The limited knowledge of the candidate storage formation can significantly influence the selection of prior geostatistical models or increase the chances of constructing prior reservoir models that are not ideal (lack of structural details compared to the ground-truth reservoir model). Secondly, injecting $CO_2$ into deep saline aquifers involves complex physics which makes it impossible to descript the process with a determined comprehensive physical simulator.

**Algorithm 2**
Data assimilation workflow to match high-dimensional observation data and generate predictions.

**Inputs**: Initial number of permeability realizations $N_r$, PCA tolerance $T$, ensemble size $N_e$, and data assimilation steps $N_a$, observation data $\mathbf{d}$.

1 Generate $N_r$ permeability fields using the fractal model in Algorithm 1.
2 Determine $n$ based on PCA tolerance $T$ and build a PCA generative model.
3 Sample $\xi_n$ from $N(0, \mathbf{I}_n)$ and $\chi$ from $U(0.05, 0.45)$ for $N_e$ times to build the initial parameter ensemble $\mathbf{m}^0 = [\mathbf{m}_1^0, \mathbf{m}_2^0, \ldots, \mathbf{m}_{N_e}^0]$.
4 Generate permeability fields $k^0 = [k_1^0, k_2^0, \ldots, k_{N_e}^0]$ from the ensemble of $\xi_n$.
5 Run saturation forward simulation and apply polar transform to get $\lambda_j$ ($j = 1, 2, \ldots, N_e$).
Run pressure forward simulation to get $\mathbf{p}_j$ and build initial predicted data ensemble $\mathbf{d}^0 = [\mathbf{d}_1^0, \mathbf{d}_2^0, \ldots, \mathbf{d}_{N_e}^0]$.
6 For $l = 1$ to $N_a$:
Obtain $\mathbf{m}^l$ based on Eq. (1) and calculate $\mathbf{d}^l$.
7 Run forward simulations to obtain $CO_2$ plume and pressure predictions in the future steps based on posterior permeability fields $k^{N_{iter}}$.

Site managers are typically not concerned with the permeability distribution and its uncertainty in the reservoir. More commonly, they need to estimate $CO_2$ plume migration and its uncertainty to prevent leakage or quantify the uncertainty of pressure magnitude to prevent induced seismicity. *A major contribution of the proposed workflow is to demonstrate the feasibility of narrowing the prediction uncertainty of these key properties of interests by efficiently assimilating monitoring data based on non-ideal prior reservoir models and imperfect physical models.*

Proposing such a workflow is a non-trivial attempt since these more than usual amount of model uncertainties (and perhaps some systematic bias) can easily lead to ensemble collapse issues. Ensemble collapse usually happens when the limited number of models in the ensemble underestimate the uncertainties (Lorentzen et al., 2019).We therefore increase the ensemble size and find that an ensemble size larger than 5000 is necessary to avoid ensemble collapse in this specific problem. The high computational efficiency of the deep learning-based surrogate models contributes to the success of the attempt. There are possibly other strategies to avoid ensemble collapse based on a smaller ensemble size, such as applying correlation-based localization techniques (Lorentzen et al., 2019; Luo and Bhakta, 2020) or implementing more advanced ensemble-based methods (Raanes et al., 2019).

### 6.2. Workflow generalizability and limitation

Although the workflow is developed based on a particular reference reservoir model, it is generalizable to other scenarios as discussed below.

The workflow only uses generic prior information including reservoir dimension, a rough range of permeability, a porosity-permeability correlation and relative permeability curves. If the workflow is to be applied for a different depositional environment, such prior information can be easily updated. Note that the fractal geostatistical model does not use any information about the two-facies structure of the multi-layer reservoir model. However, the fractal geostatistical model can still generate heterogenous permeability distributions to match apparent responses ($CO_2$ plume and pressure magnitude) from two-facies reservoir models. In cases where the fractal model fails to generate certain responses of a particular porosity/permeability structure, more prior geostatistical information should be incorporated, but the overall workflow remains the same.

The single-layer 3D reservoir model is generalizable to reservoirs that are not entirely flat, such as formations with anticlinal structures. Spatial variations of depths can be easily incorporated in the single-layer reservoir model, which will cause preferential lateral migration of the $CO_2$ plume. Because such preferences are inherently embodied in all simulation results, they can be readily "learned" by the machine-learning model.

However, if observation data and metrics to forecast include vertically resolved features, such as vertical distribution of $CO_2$ plumes, the underlying reservoir models will need to be able to resolve these features, and the surrogate models will need to be compatible with the reservoir models. For such cases, the overall structure of the workflow remains the same while developing techniques to efficiently reduce the parameter space and to build surrogate models for large-scale 3D reservoir is an ongoing endeavor.

The envisioned application scenario is that $CO_2$ plumes can be obtained from 4D or time-lapse 3D seismic inversion. In the current study, we used a rudimentary method to obtain $CO_2$ plume shapes from reservoir simulations. A major challenge associated with real-world seismic inversion is that the uncertainty in the inverted saturation is complicated and difficult to quantify. The scope of this work does not include addressing the complexity of 4D seismic data inversion, which is a significant limitation of the proposed workflow. A comprehensive summary of the challenges associated with 4D seismic data integration is available in a recent review paper by Oliver et al. (2021).

## 7. Conclusions

In this study, we developed a computationally efficient data assimilation and forecasting workflow for commercial-scale geologic carbon storage in deep saline formations. An ensemble-based framework, ES-MDA is employed to drive the data assimilation and quantify the prediction uncertainties. We also developed various techniques to efficiently reduce the dimensionality of parameter and data spaces. The computational speed of the workflow is boosted by two deep learning-based surrogate models for pressure and $CO_2$ plume predictions, respectively. We proposed a two-step approach to deal with the complex boundary conditions in multi-well injection scenarios. The surrogate models are demonstrated to be accurate and generalizable to varied permeability maps and injection durations.

In this workflow, a single-layer 3D reservoir model is employed as an intermediate link to assimilate observation data (lateral $CO_2$ plumes and monitoring pressure) generated from a high-fidelity multi-layer 3D reservoir model and to forecast future reservoir performance. This strategy is appropriate for thin 3D reservoirs where there are no practical ways to resolve the vertical features. The approach also proves effective as observation data from the high-fidelity model can be well matched and uncertainties of reservoir forecasts can be reduced based on the calibrated reservoir models.

## CRediT authorship contribution statement

**Hewei Tang:** Methodology, Software, Writing – original draft, Conceptualization. **Pengcheng Fu:** Methodology, Writing – original draft, Conceptualization. **Christopher S. Sherman:** Methodology, Software, Writing – review & editing. **Jize Zhang:** Methodology, Software, Writing – review & editing. **Xin Ju:** Software, Writing – review & editing. **François Hamon:** Software, Writing – review & editing. **Nicholas A. Azzolina:** Conceptualization, Project administration, Writing – review & editing. **Matthew Burton-Kelly:** Software. **Joseph P. Morris:** Conceptualization, Project administration, Writing – review & editing.

## Declaration of Competing Interest

The authors declare that they have no known competing financial interests or personal relationships that could have appeared to influence the work reported in this paper.

## Acknowledgments

This manuscript has been authored by Lawrence Livermore National Security, LLC under Contract No. DE-AC52–07NA2 7344 with the US. Department of Energy (DOE). The United States Government retains, and the publisher, by accepting the article for publication, acknowledges that the United States Government retains a non-exclusive, paid-up, irrevocable, world-wide license to publish or reproduce the published form of this manuscript, or allow others to do so, for United States Government purposes. This report is LLNL-JRNL-822252. This work was completed as part of the Science-informed Machine learning to Accelerate Real Time decision making for Carbon Storage (SMART-CS) Initiative (edx.netl.doe.gov/SMART). Support for this initiative came from the U.S. DOE Office of Fossil Energy's Carbon Storage Research program. The authors wish to acknowledge Mark McKoy (NETL, Carbon Storage Technology Manager), Darin Damiani (DOE Office of Fossil Energy, Carbon Storage Program Manager), and Mark Ackiewicz (DOE Office of Fossil Energy, Director, Division of Carbon Capture and Storage Research and Development), for programmatic guidance, direction, and support. We thank Dr. Grant Bromhal, Dr. Fred Aminzadeh, Dr. Srikanta Mishra, Dr. George Guthrie, Dr. Thomas McGuire, and Dr. Catherine Yonkofski for advice and support. Funding for GEOSX development was provided by Total S.A. through the FC-MAELSTROM project and the U.S. Department of Energy, Office of Science, Exascale Computing Project. FH completed this work during a visiting scientist appointment at LLNL. The codes of the workflow are available at https://github.com/tang39/DLADA.